\documentclass[prd,twocolumn,10pt,nofootinbib]{revtex4-1}

\usepackage{color}      
\usepackage[colorlinks,
            linkcolor=red,
            anchorcolor=blue,
            citecolor=cyan,     
            ]{hyperref}
\usepackage{amsmath}
\usepackage{cancel}
\usepackage{graphicx}
\usepackage{float} 
\usepackage{subfigure}
\usepackage{mathrsfs}
\usepackage{amsthm}
\usepackage{amssymb}
\usepackage{soul}
\usepackage{ulem}
\usepackage{bbm}
\usepackage{booktabs}
\usepackage{multirow}
\allowdisplaybreaks[4]
\usepackage{graphicx}
\usepackage{listings} 

\newcommand{\eref}[1]{Eq.~(\ref{#1})}

\begin{document}
	\title{Subdominant Modes of Scalar Superradiant Instability and Gravitational Wave Beats}
	
	\author{Yin-da Guo}
	\email{yinda.guo@mail.sdu.edu.cn}
	\author{Shou-shan Bao}
	\email{ssbao@sdu.edu.cn}
	\author{Hong Zhang}
	\email{hong.zhang@sdu.edu.cn}
	\affiliation{Key Laboratory of Particle Physics and Particle Irradiation (MOE),\\
		 Institute of Frontier and Interdisciplinary Science, \\
		 Shandong University, Qingdao, Shandong 266237, China}
		
	\date{\today}
	\begin{abstract}
Ultralight scalars could extract energy and angular momentum from a Kerr black hole (BH) because of superradiant instability. Multiple modes labelled with $nlm$ grow while rotating around the BH, emitting continuous gravitational waves (GWs). In this work, we carefully study the contribution of the subdominant modes with $n\geq 1$ in the evolution of the BH-condensate system. We find that the BH still evolves along the Regge trajectory of the $n=0$ modes even with the presence of the subdominant modes. The interference of the dominant and the subdominant modes produces beats in the emitted GWs, which could be used to distinguish the BH-condensate systems from other monochromatic GW sources, such as neutron stars. 
\end{abstract}
	
	\maketitle
	

\section{Introduction}\label{Introduction}

It is exciting that the gravitational wave (GW) GW150914 \cite{LIGOScientific:2016aoc} was reported by the Laser Interferometer Gravitational-Wave Observatory (LIGO) and Virgo in 2015, which demonstrates the existence and merging of a binary stellar-mass black hole (BH) system for the first time. Since then, a new window has been opened to observe the Universe. It was shown in Ref.~\cite{Barausse:2014tra} that GW astrophysics can become a precision discipline without being spoiled by various astrophysical environmental effects.

\begin{figure}[hbpt]
	\includegraphics[width=0.45\textwidth]{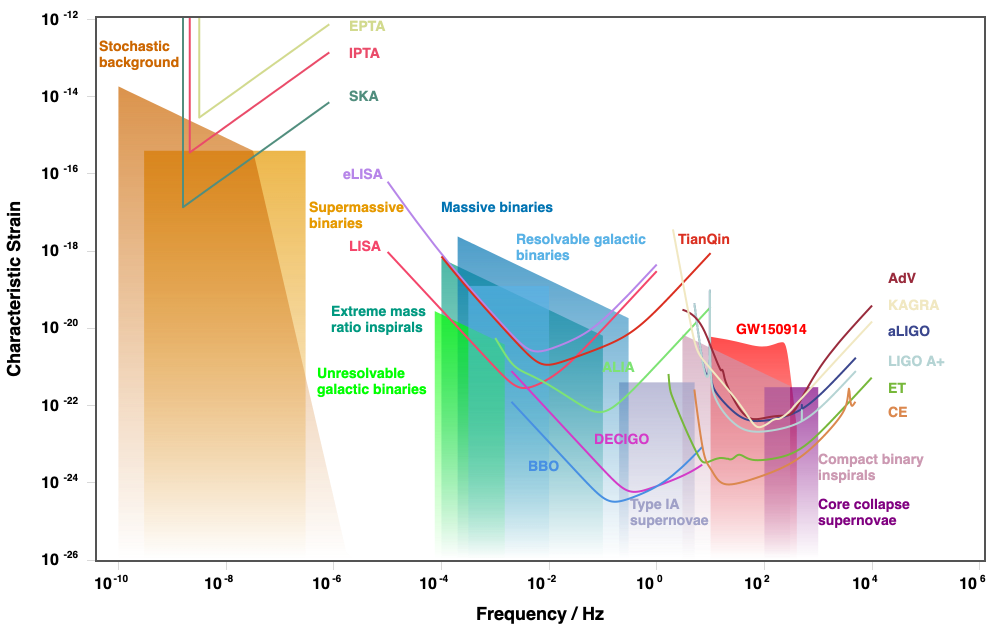}
	\caption{The various detectors' noise characteristic strains and the characteristic strains of different sources, against the GW frequency in the detector frame.  The sources of core-collapse supernovae are assumed to be 300 kpc away from us, while the sources of type IA supernovae correspond to an optimal orientation at 30 kpc. The masses of massive binaries and supermassive binaries are about $10^6$ and $10^9$ solar masses in the figure, respectively. For clarity, the resonance spikes are removed. The figure is generated by the site \url{http://gwplotter.com}. More details can be found in Ref.~\cite{Moore:2014lga}.  }
	\label{fig:_GW_Detectors_and_Sources}
\end{figure}

Many astrophysical and cosmological subjects could be studied with their special GW signals. In Fig.~\ref{fig:_GW_Detectors_and_Sources}, some important GW sources are shown, together with the sensitivity of several projected GW detectors. These detectors are divided into three categories: ground-based detectors \cite{Harry:2010zz, VIRGO:2014yos, Somiya:2011np, Hild:2010id}, space-based detectors \cite{LISA:2017pwj, TianQin:2015yph, Luo:2021Manual, Kawamura:2006up}, and pulsar timing arrays \cite{Kramer:2013kea, Manchester:2013ndt, Dewdney:2009Manual}, corresponding to GWs of high frequency, low frequency, and very low frequency, respectively. The GW sources could be binaries (including BH binaries, neutron star binaries, and white-dwarf binaries), rotating neutron stars, the stochastic background of supermassive binaries, supernovae, inflation, and so on. 

BH-condensate systems could also emit GWs. Bosonic fields centered with a Kerr BH can form bound states like the hydrogen atom. Especially, the bosonic cloud could gain sufficiently large mass and angular momentum by a superradiance mechanism such that their spacetime disturbance can produce observable GWs. There exist numerous research works on various bosons, including spin-0 \cite{Zouros:1979iw,Detweiler:1980uk,Cardoso:2005vk,Konoplya:2006br,Dolan:2007mj,Arvanitaki:2009fg,Arvanitaki:2010sy,Konoplya:2011qq,Witek:2012tr,Yoshino:2013ofa,Arvanitaki:2014wva,Brito:2014wla,Yoshino:2015nsa,Arvanitaki:2016qwi,Endlich:2016jgc,Brito:2017zvb,Brito:2017wnc,Ficarra:2018rfu,Roy:2021uye,Bao:2022hew,Chen:2022nbb,Hui:2022sri,Yuan:2022nmu}, 
spin-1 \cite{Witek:2012tr,Pani:2012vp,Pani:2012bp,Endlich:2016jgc,Baryakhtar:2017ngi,Dolan:2018dqv,East:2017mrj,East:2017ovw,East:2018glu,Frolov:2018ezx,Siemonsen:2019ebd,Percival:2020skc,Caputo:2021efm,East:2022ppo} and spin-2 \cite{Brito:2013wya,Brito:2020lup} fields. It has been shown that the superradiant process happens when the frequency $\omega$ of a bosonic wave satisfies $0<\text{Re}(\omega)<m\Omega_H$, where $m$ is the azimuthal index of the instability mode and $\Omega_H$ is the horizon angular velocity. From conservation laws, the energy and the angular momentum are extracted from the rotating BH at the center. We refer interested readers to Ref.~\cite{Brito:2015oca} for comprehensive discussions on superradiance.

The superradiant instability of such a BH-condensate system is strongest when the boson's Compton wavelength is comparable to the BH radius---i.e., $GM\mu/(\hbar c) \sim \mathcal{O}(1)$---where $M$ and $\mu$ are the masses of the BH and the boson, respectively. The Standard Model particles are all too heavy, while many ultra-light beyond-Standard Model particles, such as the QCD axion, axion-like particles suggested by string theory, and dark photons, could generate a large superradiant instability around BHs with stellar mass or heavier \cite{Arvanitaki:2009fg,Essig:2013lka,Marsh:2015xka,Hui:2016ltb,Barack:2018yly}. These bosons are also considered to be candidates for dark matter. Thus, the studies of BH superradiance and their GW signals provide an independent strategy to restrict the mass range of these dark matter candidates.

Below, we focus on the superradiant instability with ultra-light scalars. Such BH-condensate systems have been widely studied for constraining the scalar properties and for the possible observation of the GW emission. It has been shown that the BH evolves along the Regge trajectories on the mass-spin plot if the superradiant effect is strong \cite{Arvanitaki:2010sy,Brito:2014wla}. Consequently, there are ``holes" on the Regge plot in which BHs cannot reside. Combing with the observed BH merger events, favored and unfavored scalar mass ranges can be identified \cite{Ng:2019jsx,Ng:2020ruv,Cheng:2022ula}. On the other hand, with the continuous GWs generated by the BH-condensate, work has been done to study the possibility of resolving these systems from their backgrounds \cite{Arvanitaki:2010sy,Yoshino:2013ofa,Yoshino:2014wwa,Arvanitaki:2014wva,Arvanitaki:2016qwi,Brito:2017wnc,Brito:2017zvb}. One difficulty is distinguishing them from other monochromatic GW sources, such as neutron stars. In Refs.~\cite{Arvanitaki:2014wva,Baryakhtar:2017ngi}, it is proposed that the GWs from the BH-condensate systems have a positive frequency drift due to self-gravity, while the GWs from neutron stars have a frequency drift in the opposite direction. The unresolved BH-condensate systems have also been carefully studied as stochastic backgrounds for GW detectors \cite{Brito:2017wnc,Brito:2017zvb}.

The scalar condensate consists of different modes, which are usually labelled by $\{n,l,m\}$ in literature. Previous works mainly focus on the dominant modes with $l=m$ and $n=0$. In this work, we show that the subdominant modes with $l=m$ and $n>0$ also have important contributions. The evolution of the BH-condensate systems is much more complicated with the subdominant modes. Nonetheless, we find that the BH still evolves along the Regge trajectory of the $n=0$ modes even with the presence of the subdominant modes. Moreover, the life of each mode can be split into different phases: {\it accelerating, decelerating, attractor} and a possible {\it quasi-normal} phase depending on the accretion efficiency. With the observation of these phases, we find simple formulas which estimate the maximum mass and lifetime of each mode reasonably well if accretion is absent. Compared to the numerical results, these approximations work reasonably well, supporting our strategy to split the life of a mode into the phases above. It also provides a simple way to analyze the BH-condensate system without solving the differential equations. 

During the evolution, the subdominant modes could coexist with the dominant modes for a very long time. The mass ratio of the former to the latter could reach several percent. Due to the small difference in the periods,  constructive interference happens when the dense regions of two modes overlap. On the other hand,  destructive interference happens when two modes are out of phase. This results in a periodic modulation of the GW emission flux. The period of this modulation in the source frame can be estimated using Eq.~\eqref{eq:omega_R} below. It is within the scope of the current and projected GW detectors. This modulation provides another signature to break the degeneracy of the GWs emitted by some BH-condensate systems and by other monochromatic GW sources. In this work, we calculate the GW emission flux with the $\{0,1,1\}$ and $\{1,1,1\}$ modes. The interference term is suppressed by $\sqrt{N_{111}/N_{011}}$, with $N_{nlm}$ being the number of scalars in the $\{n,l,m\}$ mode. Thus, the modulation is $\sim 10\%$ even when the mass of the subdominant mode is only one percent of the dominant one. We further calculate the GW strains of the modulated waveform. We find that DECIGO has a very good potential for scalars with a mass between $10^{-16}$ and $10^{-13}$~eV. Advanced LIGO is sensitive to scalars with masses from $10^{-13}$ to $10^{-11}$~eV. LISA, Taiji, and TianQin are capable of analyzing the mass range between $10^{-18}$ and $10^{-16}$~eV.

This paper is organized as follows. In Sec. \ref{sec:_ultra-light_scalar_field_and_superradiance_modes}, we briefly review the superradiance with scalars and the calculation of instability rates for different modes. In Sec.~\ref{sec:BH_evo}, we solve the evolution of the BH-condensate system with subdominant modes, first without accretion, and then with the effect of the accretion and nonlinearity argued. Then, in Sec.~\ref{sec:GW_emission}, we calculate the GW emission with the interference of $\{0,1,1\}$ and $\{1,1,1\}$ modes. The beat-like modulation is quite strong and could be observed by current and projected GW telescopes. Finally, we summarize in Sec.~\ref{sec:_Summary}.

\section{Light scalar in Kerr metric}
\label{sec:_ultra-light_scalar_field_and_superradiance_modes}

In Boyer-Lindquist coordinates, the Kerr metric  with mass $M$ and angular momentum $J$ can be expressed as \cite{Boyer:1966qh},
\begin{equation}
	\begin{aligned}
	d s^{2}=&\left(1-\frac{2 M r}{\Sigma}\right) d t^{2}+\frac{4 a M r}{\Sigma} \sin ^{2} \theta d t d \varphi-\frac{\Sigma}{\Delta} d r^{2}& \\
	&
	-\Sigma d \theta^{2}-\left[\left(r^{2}+a^{2}\right) \sin ^{2} \theta+2 \frac{M r}{\Sigma} a^{2} \sin ^{4} \theta\right] d \varphi^{2},
	\end{aligned}
\end{equation}
where, 
\begin{equation}
	\begin{aligned}
		a &=J / M, \\
		\Delta &=r^{2}-2 M r+a^{2}, \\
		\Sigma &=r^{2}+a^{2} \cos ^{2} \theta.
	\end{aligned}
\end{equation}
The inner and outer horizons $r_\pm$ are located at, 
\begin{equation}
  	r_{\pm}=M \pm \sqrt{M^{2}-a^{2}}.
\end{equation}
Here we adopt the Planck units $G=\hbar = c = 1$.

In this work, we consider a real scalar field surrounding a Kerr BH. It has been shown in Ref.~\cite{Brito:2014wla} that the backreaction of the scalar condensate is negligible because of its low energy density. It is also qualified to ignore self-interaction for the same reason. These contributions may be taken into account with the perturbation method if high-precision results are needed, which is beyond the scope of this work. Here we consider a free real scalar field on the Kerr metric,
\begin{equation}
	\left(\nabla^{\mu}\nabla_{\mu}+\mu^{2}\right)\Phi=0,
	\label{eq:_KG_equation}
\end{equation}
where $\mu$ is the mass of the scalar particle. The term with the d'Alembert operator can be written in an expanded form,
\begin{equation}
  \begin{aligned}
	\nabla^{\mu}\nabla_{\mu}\Phi=\dfrac{1}{\Sigma} & \left\{\left[\frac{\left(r^{2}+a^{2}\right)^{2}}{\Delta}-a^{2}\sin^{2}\theta\right]\frac{\partial^{2}}{\partial t^{2}}\right.\\
	&-\frac{\partial}{\partial r}\left(\Delta\frac{\partial}{\partial r}\right)-\frac{1}{\sin\theta}\frac{\partial}{\partial\theta}\left(\sin\theta\frac{\partial}{\partial\theta}\right)\qquad \\
	&+\left.\left[\frac{a^{2}}{\Delta}-\frac{1}{\sin^{2}\theta}\right]\frac{\partial^{2}}{\partial\varphi^{2}}+\frac{4Mar}{\Delta}\frac{\partial^{2}}{\partial t\partial\varphi}\right\}\Phi.
  \end{aligned}
\end{equation} 
The solution of Eq.~\eqref{eq:_KG_equation} can be written as,
\begin{equation}
	\Phi\left(t,r,\theta,\varphi\right)=\sum_{l,m}\int d\omega\dfrac{f\left(\omega\right)}{\sqrt{2\omega}}\left(\phi_{l m}+\phi_{l m}^{*}\right),
	\label{eq:_wave_function_1}
\end{equation}
where $\omega$ is the eigenfrequency of the scalar field and $f\left(\omega\right)$ is the distribution function which will be explained later. The variables in $\phi_{lm}$ can be further separated \cite{Teukolsky:1973ha, Press:1973zz},
\begin{equation}
	\phi_{l m}\left(t,r,\theta,\varphi\right)=e^{-i\omega t}\dfrac{e^{im\varphi}}{\sqrt{2\pi}}\mathcal R_{l m}(r)\mathcal S_{l m}(\theta).
	\label{eq:_def_phi}
\end{equation}
The functions $\mathcal{R}_{l m}(r)$ and $\mathcal S_{l m}(\theta)$ satisfy the following equations,
\begin{align}
\begin{split}\label{eq:_radial_equation_of_KG_eq}
& \Delta \frac{d}{d r}\left(\Delta \frac{d \mathcal{R}_{l m}(r)}{d r}\right)+\left[\omega^{2}\left(r^{2}+a^{2}\right)^{2}-\right. 4 a M r m \omega \\
&\hspace{0.5cm}
+a^{2} m^{2}-\Delta\left(\mu^{2} r^{2}+a^{2} \omega^{2}+\Lambda_{l m}\right) \Big] \mathcal{R}_{l m}(r)=0,
\end{split}\\
\begin{split}\label{eq:_spheroidal_harmonics}
&\frac{1}{\sin \theta} \frac{d}{d \theta}\left(\sin \theta \frac{d \mathcal{S}_{l m}(\theta)}{d \theta}\right) +\Big[-a^{2}\left(\mu^{2}-\omega^{2}\right) \cos ^{2} \theta\\
&\hspace{3.5cm}
-\frac{m^{2}}{\sin ^{2} \theta}+\Lambda_{l m}\Big] \mathcal{S}_{l m}(\theta)=0,
\end{split}
\end{align}
where $\Lambda_{l m}$ and the spheroidal harmonics $\mathcal{S}_{l m}(\theta)$ are the eigenvalue and eigenfunction of the angular equation, respectively. They can be solved numerically with Leaver's continued fraction method \cite{Leaver:1985ax,Leaver:1986Manual,Cardoso:2005vk}. They can also be obtained conveniently in \textsc{Mathematica} with the functions \verb|SpheroidalEigenvalue| and \verb|SpheroidalPS|. 

In this work, we choose the normalization of $\mathcal S_{l m}(\theta)$ and $\mathcal R_{l m}(r)$  as,
\begin{gather}
  \int^\pi_0\left|\mathcal S_{l m}(\theta)\right|^2\sin{\theta}d\theta =1,\\
\int^\infty_{r_+} \int^\pi_0 \left|\mathcal R_{l m}(r) \mathcal S_{l m}(\theta)\right|^2 \sqrt{|g|} d r d \theta  =1,
\end{gather}
where $\sqrt{|g|}=\left(r^2+a^2 \cos^2\theta\right)\sin\theta$. With these normalization conditions, $|f(\omega)|^2$  is the particle number density with frequency $\omega$ in the condensate.

For bounded scalar fields, both the numerical solutions and analytic approximations are used to solve for $\omega$ from \eref{eq:_radial_equation_of_KG_eq}. To simplify the numerical calculation, Leaver proposed a continued fraction method for massless fields \cite{Leaver:1985ax,Leaver:1986Manual}. The eigen-equation is converted into a three-term recurrence relation. Leaver's method was developed for massive scalar fields in Ref.~\cite{Cardoso:2005vk} and refined later in Ref.~\cite{Dolan:2007mj}. On the other hand, the analytic approximation at the limit $\mu M \ll 1$ is obtained in Ref.~\cite{Detweiler:1980uk}. This approximation is not consistent with the numerical solution, causing doubts about the reliability of both results. The problem is recently resolved in Ref.~\cite{Bao:2022hew} by including the next-to-leading-order (NLO) contribution in the analytic result. At the other limit $\mu M \gg 1$, a JWKB estimate of the fastest growth rate is given in Ref.~\cite{Zouros:1979iw}.  

In general, the obtained eigenvalues $\omega$ are complex numbers depending on three indices $\{n,l,m\}$. Another widely-used index $\bar{n}$ is defined as $n+l+1$. We use $\omega_R$ and $\omega_I$ to denote the real and imaginary parts of $\omega$, respectively. The superradiance condition $\omega_I>0$  requires $\omega_R<\omega_c \equiv m\Omega_H$, where $\Omega_H \equiv a/(2M r_+)$ is the angular velocity at the outer horizon.

\begin{figure}
	\includegraphics[width=0.4\textwidth]{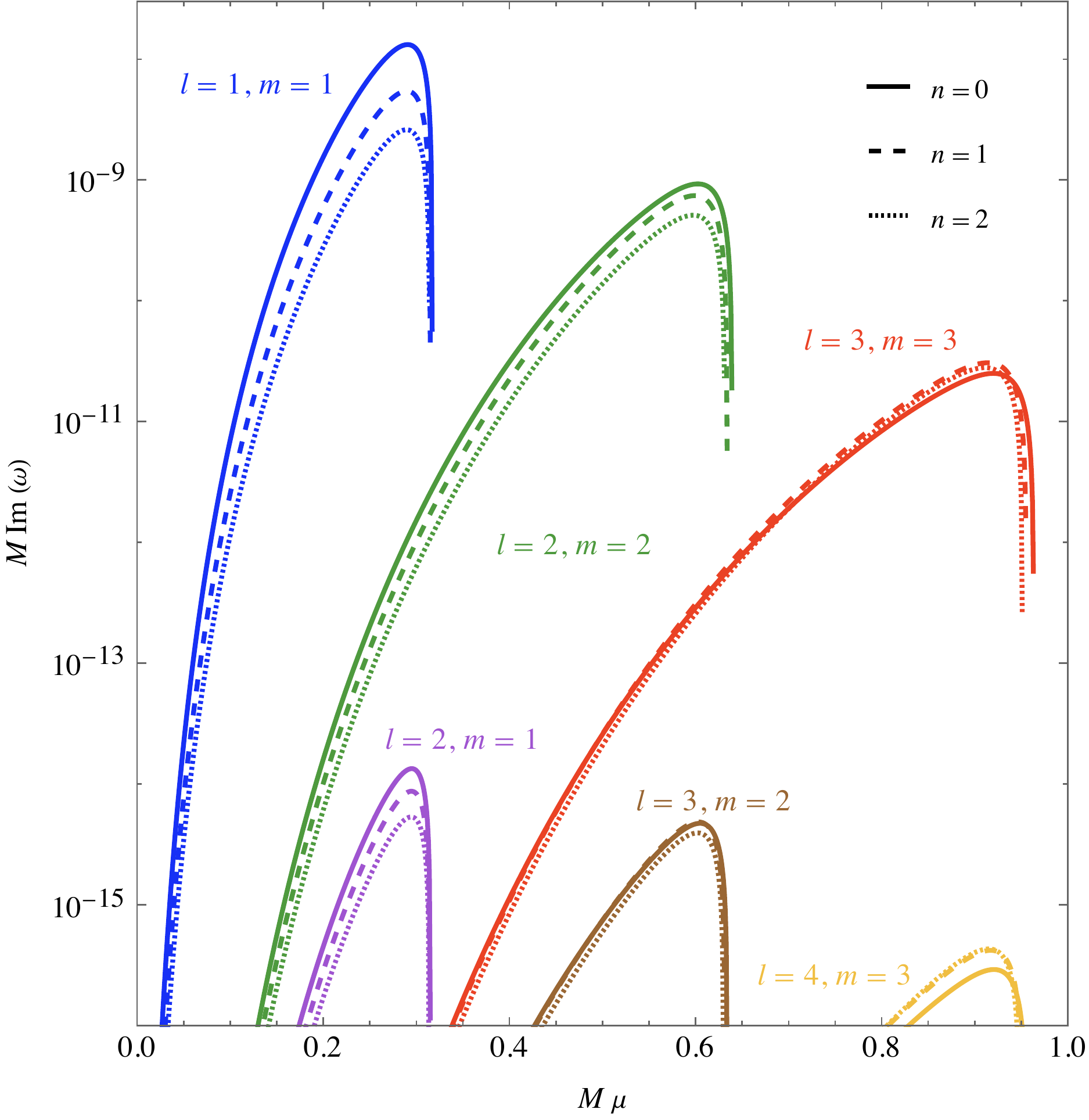}
	\caption{$\omega_I$ as a function of the multiplication of the BH mass $M$ and the real scalar mass $\mu$, calculated with the NLO analytic approximation in Ref.~\cite{Bao:2022hew}. The BH angular momentum $a_*$ is chosen to be $0.9$. The curves with different colors correspond to different values of $l$ and $m$. For each set of $l$ and $m$, three modes with $n=0$ (solid), $n=1$ (dashed), and $n=2$ (dotted) curves are shown.}
	\label{fig:_anay_Im_omega}
\end{figure}

The function $\omega_I$ has been well-studied in literature. Below, we list only some properties which are important for this work. Fig.~\ref{fig:_anay_Im_omega} shows $\omega_I$ as a function of the product of the black hole mass $M$ and the scalar mass $\mu$ for some important sets of $\{n,l,m\}$, using the NLO analytic approximation in Ref.~\cite{Bao:2022hew}. The dimensionless BH spin is fixed at $a_*\equiv a/M=0.9$. For $M\mu\lesssim 0.3$, the fastest-growing mode is $\{0,1,1\}$. In the same region of $M\mu$, the $\{n,1,1\}$ modes with $n>0$ decrease in importance with increasing value of $n$. To the right of the fast-dropping edges of $\{n,1,1\}$ modes, the $l=m=2$ modes are the most important. But their values are smaller than those with $l=m=1$. Fig.~\ref{fig:_anay_Im_omega} also shows that modes with $m<l$ grow too slowly to be important in phenomenology.

\begin{figure}
	\includegraphics[width=0.4\textwidth]{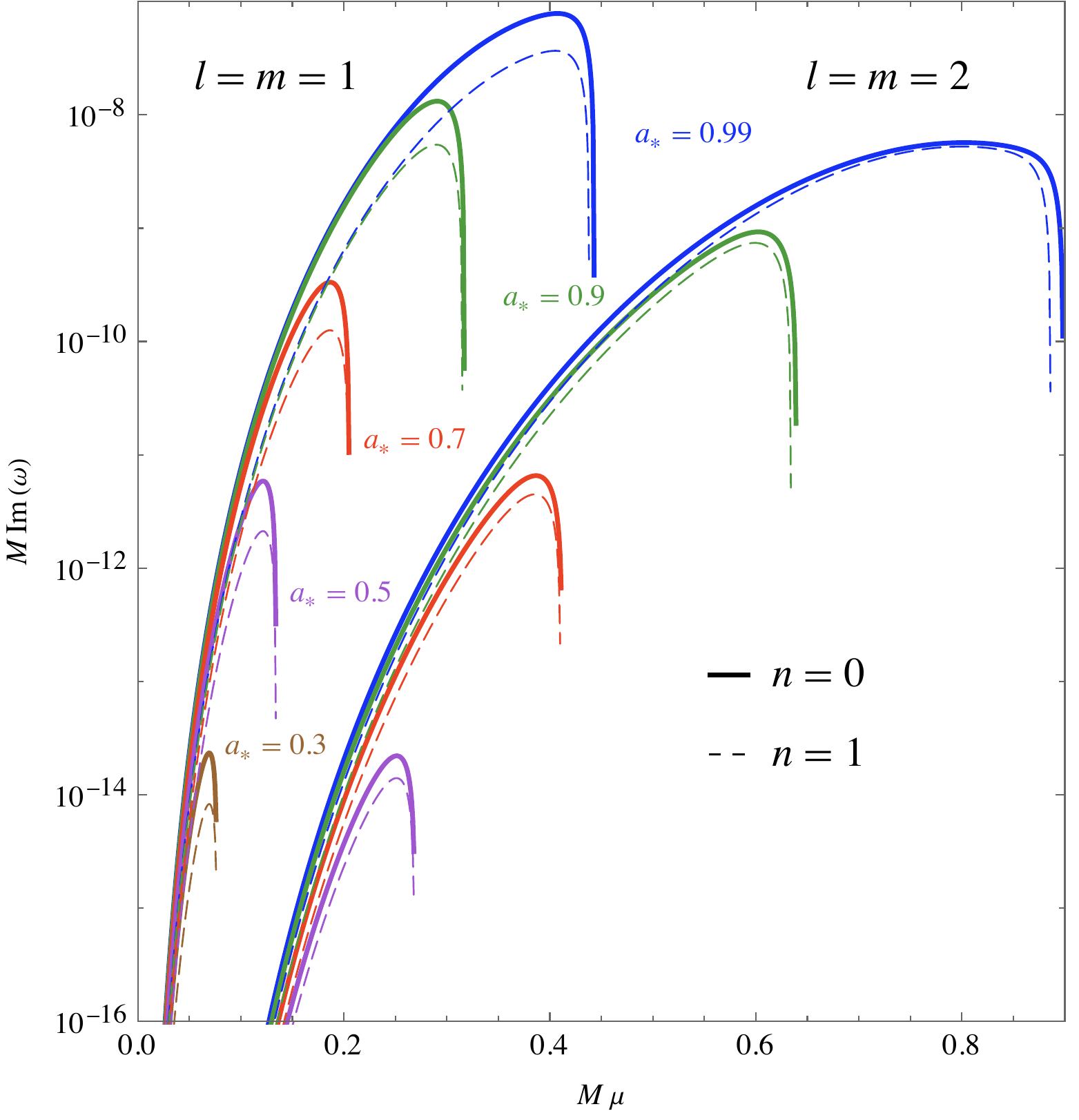}
	\caption{The function $\omega_I$ of the $l=m=1$ and $l=m=2$ modes as a function of the multiplication of the BH mass $M$ and the real scalar mass $\mu$, for different values of $a_*$. Both the $n=0$ (solid) and $n=1$ (dashed) curves are shown. The results are calculated with the NLO analytic approximation in Ref.~\cite{Bao:2022hew}. }
	\label{fig:_anay_Im_omega_a=0.99_to_0.4}
\end{figure}

To study the contribution of subdominant modes, we consider the four modes $\{0,1,1\}$, $\{1,1,1\}$, $\{0,2,2\}$, and $\{1,2,2\}$ in this work. Their $\omega_I$'s as functions of $M\mu$ are plotted with different values of $a_*$ in Fig.~\ref{fig:_anay_Im_omega_a=0.99_to_0.4}. For each $\{n,l,m\}$ mode, decreasing the value of $a_*$ leads to a lower curve with the peak moving to the left. Away from the peaks, the curves with different $a_*$ values differ very little on the logarithmic scale. While in the neighborhood of the peaks, curves with different $a_*$'s have very different behaviors.


\section{Contribution to BH-Condensate Evolution}\label{sec:BH_evo}

In this section, we study the contribution of subdominant modes to the BH-condensate evolution and the GW emission luminosity. Particularly, we focus on the limit $M\mu\lesssim 1$, in which region the use of NLO analytic approximation for $\omega_I$ is qualified \cite{Bao:2022hew}. Two dominant modes ($\{0,1,1\}$ and $\{0,2,2\}$) and two subdominant modes ($\{1,1,1\}$ and $\{1,2,2\}$) are considered. We first carefully calculate the evolution of the BH-condensate system without accretion in Sec.~\ref{subsec:withoutAcc}, then argue the effect of accretion in Sec.~\ref{subsec:withAcc}, with the help of the results in Ref.~\cite{Brito:2014wla}. Finally, we discuss  the nonlinear effects briefly in Sec.~\ref{subsec:nonlinear}.

\subsection{Without Accretion}\label{subsec:withoutAcc}

We use the quasi-adiabatic approximation, since the timescales of the superradiant instability and the GW emission are much longer than the dynamical timescale of the BH \cite{Brito:2014wla,Brito:2017zvb}. Without accretion, the evolution equations are,
\begin{subequations}\label{eq:Evoluton_ODE}
\begin{align}
&	\dot{M}  = - \sum_{nlm} 2 M_{s}^{(nlm)} \omega_{I}^{(nlm)}, \\
&	\dot{J}  = - \sum_{nlm} 2 m M^{(nlm)}_{s} \omega_{I}^{(nlm)} /\omega_R^{(nlm)}, \\
&	\dot{M}_s^{(nlm)}   =  2 M_{s}^{(nlm)} \omega_{I}^{(nlm)} -  \dot{E}_\mathrm{GW}^{(nlm)}, \label{eq:Etot}\\
&	\dot{J}_{s}^{(nlm)}  = 2 m M^{(nlm)}_{s} \omega_{I}^{(nlm)} /\omega_R^{(nlm)} -  m \dot{E}_\mathrm{GW}^{(nlm)}/\omega_R^{(nlm)}
	\label{eq:Jtot},
\end{align}
\end{subequations}
where $M_s^{(nlm)}$ and $J_s^{(nlm)}$ are the mass and angular momentum of the scalar condensate with the indices $\{n,l,m\}$, respectively. $\dot{E}_\mathrm{GW}^{(nlm)}$ is the GW emission luminosity from the $\{n,l,m\}$ mode. Note that the interference between different modes is not considered, which includes transition as well as the annihilation of two scalars in different modes. The GW emitted by the former is small \cite{Arvanitaki:2010sy, Arvanitaki:2014wva}, while the latter could cause $\sim 10\%$ variation of GW luminosity and beat-like waveforms, which is the topic of Sec.~\ref{sec:GW_emission}. An approximation of $\dot{E}_\mathrm{GW}^{(011)}$ in the limit $M\mu \ll 1$ is obtained in Ref.~\cite{Brito:2014wla} with a Schwarzschild background metric. Following the same method, we obtain the results for the other three modes considered in this work. The results are listed below,
\begin{subequations}\label{eq:GW-rad}
\begin{align}
\dot{E}^{(011)}_{\mathrm{GW}}&=\frac{484+9 \pi^{2}}{23040}\Big(\frac{M_{s}^{(011)}}{M}\Big)^2(M \mu)^{14},\\
	\dot{E}^{(111)}_{\mathrm{GW}}&=\frac{128\left(484+9 \pi^{2}\right)}{23914845}\Big(\frac{M_{s}^{(111)}}{M}\Big)^2(M \mu)^{14}\label{eq:_dE/dt_111_approx},\\
	\dot{E}^{(022)}_{\mathrm{GW}}&=\frac{\left.1024+49 \pi ^2\right. }{5423886846}\Big(\frac{M_{s}^{(022)}}{M}\Big)^2(M \mu)^{18} \label{eq:_dE/dt_022_approx},\\
	\dot{E}^{(122)}_{\mathrm{GW}}&=\frac{\left.1024+49 \pi ^2\right. }{15032385536}\Big(\frac{M_{s}^{(122)}}{M}\Big)^2(M \mu)^{18} \label{eq:_dE/dt_122_approx}.
\end{align}
\end{subequations}

\begin{figure}
\hspace{-0.5cm}
        \includegraphics[width=0.5\textwidth]{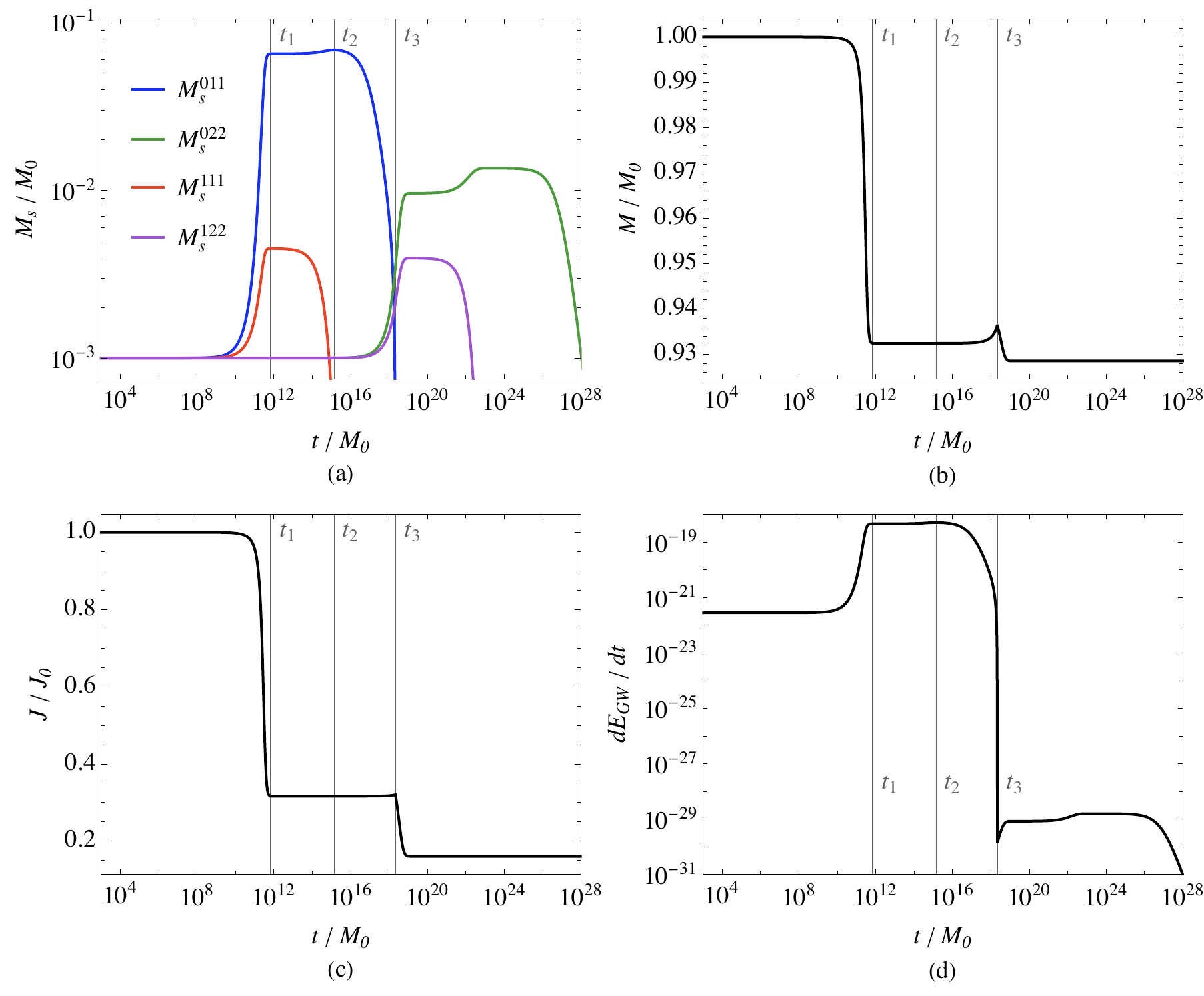}
	\caption{The evolution of the scalar condensate masses, BH mass, BH spin, and the GW emission luminosity as a function of time. The initial parameters are $M_0\mu=0.1$ and $a_*=0.99$, with $M_0$ being the initial BH mass. The initial masses of the scalar condensate in different modes are all chosen as $10^{-3}M_0$. The three vertical lines label the three critical times (see text for details). The effects of all four modes are included for each curve in panels (b), (c), and (d).
	}
	\label{fig:_evolution-0.1-0.99}
\end{figure}

We solve Eqs.~\eqref{eq:Evoluton_ODE} with the GW emission energy flux given in Eqs.~\eqref{eq:GW-rad}, keeping two dominant modes ($\{0,1,1\}$ and $\{0,2,2\}$) as well as two subdominant modes ($\{1,1,1\}$ and $\{1,2,2\}$). We choose the initial BH mass such that $M_0\mu=0.1$, and the initial BH spin $a_{*0}=0.99$. The initial mass of each of the four modes is $10^{-3} M_0$. The results are shown in Fig.~\ref{fig:_evolution-0.1-0.99}. We have used $M_0$ to normalize the time $t$. The relation to the SI unit can be obtained with,
\begin{align}\label{eq:M_SI_unit}
M_0 = 1.56\times 10^{-13} \text{yr} \left(\frac{M_0}{M_\odot}\right).
\end{align}
There are several stages in the evolution process:
\begin{itemize}
\item 
The time period $0<t\leq t_1$, with $t_1=7.03\times 10^{11} M_0$ being the time when the $\{1,1,1\}$ mode reaches its maximum mass $4.50\times 10^{-3} M_0$. In this stage, the $\{0,1,1\}$ mode rises the fastest, followed by the $\{1,1,1\}$ mode. The other two modes also increase, but the effect is too small to be visible in Fig.~\ref{fig:_evolution-0.1-0.99}. At $t_1$, the $\{0,1,1\}$ mode has mass $6.51\times 10^{-2} M_0$, the BH mass is $0.932 M_0$, and the BH spin is $0.316J_0$. The integrated GW emission energy in this stage is $1.87\times 10^{-7}M_0$.
\item 
The time period $t_1<t\leq t_2$, with $t_2=1.45\times 10^{15} M_0$ being the time when the $\{0,1,1\}$ mode reaches its maximum mass $6.86\times 10^{-2} M_0$. In this stage, the $\{1,1,1\}$ mode shrinks and is nearly depleted at $t_2$. The dominant mode $\{0,1,1\}$ rises at a much slower rate, forming a plateau in Fig.~\ref{fig:_evolution-0.1-0.99}. The BH mass and spin are approximately unchanged, with the values $0.932 M_0$ and  $0.316J_0$, respectively. The integrated GW emission energy in this stage is $7.16\times 10^{-4}M_0$.
\item 
The time period  $t_2<t\leq t_3$, with $t_3=2.18\times 10^{18} M_0$ being the time when the BH mass reaches its local maximum $0.936 M_0$. In this stage, the $\{0,1,1\}$ mode shrinks and the $l=m=2$ modes rise quickly on the logarithmic scale. At $t_3$, the BH spin also rises slightly to $0.320J_0$. The integrated GW emission energy in this stage is $6.13\times 10^{-2}M_0$.
\end{itemize}
After $t_3$, the contraction of the $\{0,1,1\}$ mode is not enough to support the growth of the $l=m=2$ modes, which then govern the evolution afterward. The three stages above repeat for the $l=m=2$ modes. The maximum mass of the $\{0,2,2\}$ mode is $1.35\times 10^{-2}M_0$. The BH mass and angular momentum further drop to $0.928 M_0$ and $0.160 J_0$, respectively. Note that the maximum mass of the $\{0,2,2\}$ mode is smaller than that of $\{0,1,1\}$ mode by a factor of 5.07, while its maximum GW emission luminosity is smaller by a factor of $\sim 10^{10}$.

The evolution of different scalar clouds can be understood with Fig.~\ref{fig:_anay_Im_omega_a=0.99_to_0.4}. Since the BH mass varies very little during the whole process, the product $M\mu$ could be taken as the constant $0.1$ approximately. Initially, the BH spin $a_*$ is $0.99$ and all four modes have $\omega_I>0$. The $l=m=1$ modes are more important, since the $\omega_I$ values for $l=m=2$ modes are several orders of magnitude smaller. With the BH angular momentum extracted by the scalar condensates, the value of $a_*$ decreases, and the $\omega_I$'s of all modes drop accordingly. The $\{1,1,1\}$ mode first reaches the critical superradiance condition $\omega_R^{111} = \Omega_H$ at time $t_1$ and starts to shrink afterwards. Until the $\{1,1,1\}$ mode is depleted, the mass and angular momentum of the BH stay almost unchanged. Mathematically, it is an attractor of the BH mass and spin, which is referred to as the $\{1,1,1\}$ attractor below. This attractor disappears as soon as the $\{1,1,1\}$ mode is nearly depleted at $t_2$ when the dominant mode $\{0,1,1\}$ has its maximum mass. Soon after $t_2$, the $\{0,1,1\}$ mode reaches its critical superradiance condition. It is the second attractor of the system, which will be called the $\{0,1,1\}$ attractor. If there were no $l=m=2$ modes or GW radiation, this attractor would have an infinite lifetime and all observables such as the BH mass and spin would not change any longer. In fact, with the BH spin being extracted by the $l=m=2$ modes and radiated by the GW, the $\{0,1,1\}$ mode shrinks and returns the mass and angular momentum to the BH. Finally, the $\{0,1,1\}$ mode is drained, and the later evolution is governed by the $l=m=2$ modes. This later evolution qualitatively repeats the previous three stages, with a much slower pace, due to the much smaller $\omega_I$ of the $l=m=2$ modes. 

The fact that larger $n$ modes reach the critical superradiance condition earlier can also be understood quantitatively at the $M\mu\ll 1$ limit. The asymptotic expression of $\omega_R$ is, 
\begin{align}\label{eq:omega_R}
\omega_R^{nlm}= \mu\left(1-\frac{(M\mu)^2}{2\bar{n}^2}\right),
\end{align} 
where $\bar{n}=n+l+1$. Here, $\omega_R$ is a monotonically increasing function of $n$. Combining with the superradiance condition, the critical BH spin $a^{nlm}_{*C}$ for mode $\{n,l,m\}$ can be obtained,
\begin{align}\label{eq:aStarC}
a^{nlm}_{*C} = \frac{4m M \omega_R^{nlm}}{m^2+(2 M \omega_R^{nlm})^2}.
\end{align}
The superradiant instability happens only at $M\omega_R^{nlm}< m/2$ (see Fig.~\ref{fig:_anay_Im_omega}), in which region $a^{nlm}_{*C}$ is an increasing function of $\omega_R^{nlm}$, and consequently an increasing function of $n$. Therefore, fixing $l$ and $m$, the mode with a larger value of $n$ reaches the superradiance condition earlier. One can further conclude that the $\{n,l,m\}$ modes with $n\geq 2$ are less important since they have a smaller $\omega_I$ and a shorter growing time. A caveat is that one could always make the large $n$ modes dominant by setting their initial masses to be several orders of magnitude larger than the small $n$ modes. This initial condition may happen if a BH passes by a scalar-dense area. We do not consider this possibility in this work. A relevant discussion of the dependence on the initial condensate mass can be found in Ref.~\cite{Ficarra:2018rfu}.

\begin{figure}
	\centering
 \includegraphics[width=0.4\textwidth]{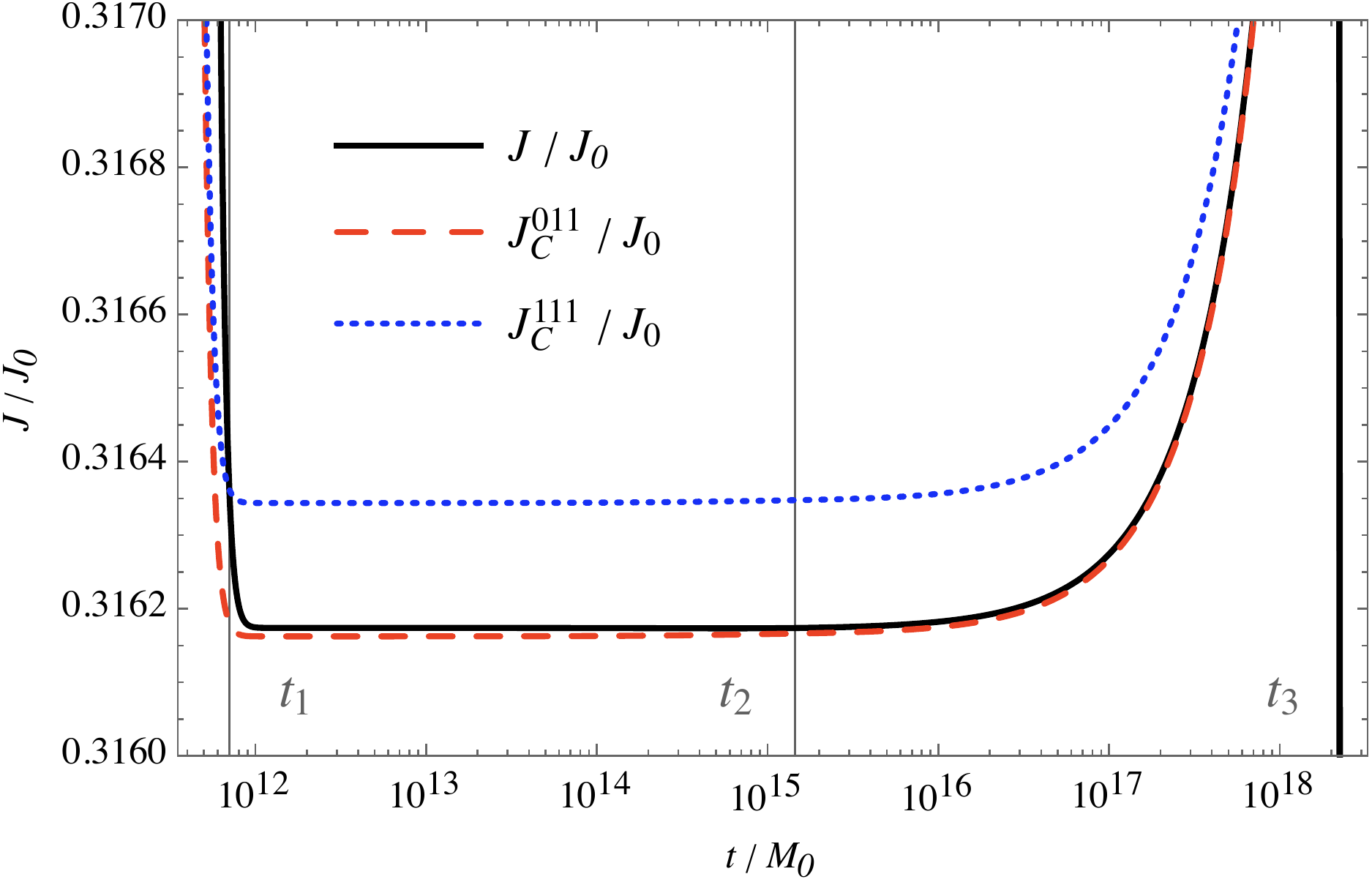}
	\caption{Comparison of the $\{1,1,1\}$ attractor (black solid curve) to the $\{0,1,1\}$ Regge trajectory (red dashed) and the $\{1,1,1\}$ Regge trajectory (blue dotted). The two Regge trajectories are calculated with $J=a_{*C}^{011} M^2$ and $J=a_{*C}^{111} M^2$.
	}
	\label{fig:Attractors}
\end{figure}

An additional explanation of the two attractors has to be made. The $\{0,1,1\}$ attractor has been well-discussed in literature \cite{Benone:2014ssa,Brito:2014wla,Hui:2022sri}. With this attractor, the BH tightly follows the $\{0,1,1\}$ Regge trajectory, determined by $J=a_{*C}^{011} M^2$ in the mass-spin plot. To the contrary, the $\{1,1,1\}$ attractor is approximately given by,
\begin{align}\label{eq:111_attractor}
M_s^{(011)} \omega_I^{(011)} + M_s^{(111)} \omega_I^{(111)}  = 0.
\end{align}
Since $M_s^{(011)}$ is much larger than $M_s^{(111)}$, the BH does not follow $J=a_{*C}^{111}M^2$ at the $\{1,1,1\}$ attractor. In Fig.~\ref{fig:Attractors}, we show the comparisons of the BH spin with the two Regge trajectories calculated with Eq.~\eqref{eq:aStarC}. The $\{1,1,1\}$ attractor is actually closer to the $\{0,1,1\}$ Regge trajectory than to the $\{1,1,1\}$ Regge trajectory in the time range $(t_1, t_2)$. Thus, it is quite accurate to state that the BH follows the $\{0,1,1\}$ Regge trajectory since time $t_1$. This statement is still valid if more subdominant modes are included in the calculation.

\begin{figure}
	\centering
 \includegraphics[width=0.5\textwidth]{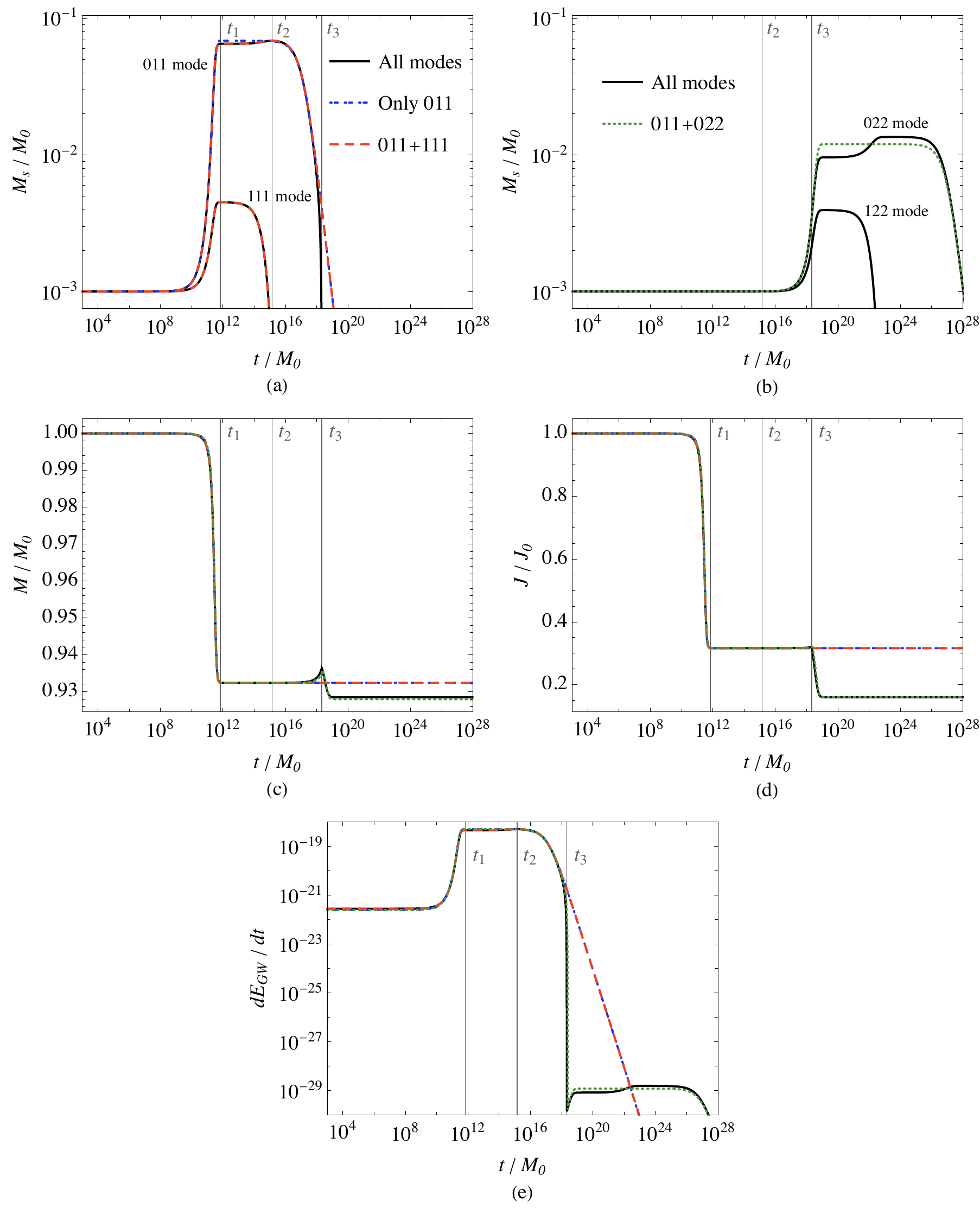}
	\caption{The evolution of the scalar condensate masses, BH mass, BH spin, and the GW emission luminosity as a function of time, keeping different combinations of subdominant modes. The initial parameters are $M_0\mu=0.1$ and $a_{*0}=0.99$, with $M_0$ being the initial BH mass. The initial masses of the scalar condensate in different modes are all chosen as $10^{-3}M_0$.  The three vertical lines label the three critical times (see text for details). In panels (c), (d), and (e), the solid (black), dashed (red), dotted (green), and dot-dashed (blue) curves are the results with all four modes, with the $\{0,1,1\}+\{1,1,1\}$ modes, with the $\{0,1,1\}+\{0,2,2\}$ modes, and with only the $\{0,1,1\}$ mode, respectively. 
	}
	\label{fig:_evolution-0.1-0.99-diff-scenarios}
\end{figure}

To investigate the contribution of different modes, we consider three more scenarios:  with the $\{0,1,1\} + \{0,2,2\}$ modes, with the $\{0,1,1\} + \{1,1,1\}$ modes, and with only the $\{0,1,1\}$ mode. The results are shown in Fig.~\ref{fig:_evolution-0.1-0.99-diff-scenarios}, in which the plot of the scalar condensate masses is split into two panels for clarity. With all the calculations above, several observations can be made.
\begin{enumerate}
\item 
The presence of the subdominant modes changes the maximum mass of the dominant modes only at the percentage level.
\item
With the same $l$ and $m$, the masses of the dominant mode and the subdominant mode reach the plateaus almost simultaneously. 
\item
BH mass and spin are approximately constants in the two attractor phases of the $l=m=1$ and $l=m=2$ modes. The dimensionless spin is very close to $a_{*C}^{011}$ and $a_{*C}^{022}$ in the two attractor phases, respectively.
\item
The $l=m=2$ modes are negligible at $t<t_3$.
\item
From the two observations above, almost all the energy and angular momentum of the $\{0,1,1\}$ mode are dissipated by GW emission.
\item
From Eq.~\eqref{eq:111_attractor}, almost all the energy and angular momentum of the $\{1,1,1\}$ mode are absorbed by the $\{0,1,1\}$ mode.
\end{enumerate}
With these observations, one could make quite accurate estimates of the important quantities without solving the differential equations, at least at the $M\mu\ll 1$ limit.

We first calculate $M_s^{(011)}$ at time $t_1$. In the time range $t<t_1$, one could only consider the BH and the $\{0,1,1\}$ mode without GW emission. From energy and angular momentum conservation, there are,
\begin{subequations}\label{eq:estimate_Ms011}
\begin{align}
M_0 &= M_1 + M_s^{(011)}(t_1),\\
a_{*0} M_0^2 &= a_{*C}^{(011)} M_1^2 + \frac{M_s^{(011)}(t_1)}{\mu},
\end{align} 
\end{subequations}
where $M_1$ is the BH mass at $t_1$. With $a_{*C}^{(011)}$ calculated using Eq.~\eqref{eq:aStarC}, one arrives at,
\begin{align}\label{eq:ratio_M011_M0}
\frac{M_s^{(011)}(t_1)}{M_0} \approx M_0\mu \,(a_{*0}-4M_0\mu).
\end{align}
Inserting $M_0 \mu=0.1$ and $a_{*0}=0.99$, we obtain ${M_s^{(011)}}(t_1)/{M_0}=5.90\times 10^{-2}$, only slightly smaller than $6.51\times 10^{-2}$ from numerical calculation.

The maximum of $M_s^{(111)}$ could be estimated with the second observation above. At time $t_1$, the masses of the $l=m=1$ modes are roughly,
\begin{align}\label{eq:Ms_nlm}
M_s^{(nlm)}(t_1) \approx M_{s0}^{(nlm)} \exp\left(2\omega_I^{nlm} t_1\right),
\end{align} 
where $M_{s0}^{(nlm)}$ is the initial mass of the $\{n,l,m\}$ mode. The dependence of $\omega_I^{(nlm)}$ on $n$ could be easily factored out with the LO analytic expression in the $M\mu\ll 1$ limit \cite{Detweiler:1980uk,Pani:2012bp,Bao:2022hew},
\begin{align}
\omega_I^{nlm} \propto \beta_{nl} \equiv \frac{(n+2l+1)!}{n!(n+l+1)^{2l+4}}.
\end{align}
Then, one could obtain the mass ratio at $t_1$,
\begin{align}\label{eq:ratio_M111_M0}
\frac{M_s^{(111)}(t_1)}{M_0} \approx \frac{M_{s0}^{(111)}}{M_0} \left(\frac{M_{s}^{(011)}(t_1)}{M_{s0}^{(011)}}\right)^{\beta_{11}/\beta_{01}}.
\end{align}
With $M_s^{(011)}$ estimated using Eq.~\eqref{eq:ratio_M011_M0} and $M_{s0}^{(nlm)} = 10^{-3}M_0$, one could get $M_s^{(111)}(t_1)/M_0 = 4.19\times 10^{-3}$, which is only slightly smaller than the numerical value $4.50\times 10^{-3}$. Then, the mass $M_s^{(011)}$ at time $t_2$ is the addition of $M_s^{(011)}$ and $M_s^{(111)}$ at time $t_1$, which is $6.32\times 10^{-2} M_0$. For comparison, the numerical value is $6.96\times 10^{-2}M_0$.

The BH mass from $t_1$ to $t_3$ is nearly a constant. At the time $t_1$, the BH mass could be estimated with,
\begin{align}
M_1 \approx M_0 -M_s^{(011)}(t_1) -M_s^{(111)}(t_1) = 0.9368 M_0,
\end{align}
which is compared to the value $0.9324 M_0$ from solving the differential equations.

The masses of the $l=m=2$ modes can also be estimated. We define $t_4$ as the time when the $\{1,2,2\}$ mode reaches its maximum mass. For $t<t_4$, the BH has mass $M_1$ and spin $a_{*C}^{(011)}$ for most of the time. Thus, they are the ``initial" BH mass and spin for the $l=m=2$ modes. Similarly to Eqs.~\eqref{eq:estimate_Ms011}, the equations for $l=m=2$ modes are,
\begin{subequations}\label{eq:estimate_Ms022}
\begin{align}
M_1 &= M_4 + M_s^{(022)}(t_4),\\
a_{*C}^{(011)} M_1^2 &= a_{*C}^{(022)} M_4^2 + \frac{2M_s^{(022)}(t_4)}{\mu},
\end{align} 
\end{subequations}
where $M_4$ is the BH mass at $t_4$. At the $M_1\mu\ll 1$ limit, one obtains,
\begin{align}
\frac{M_s^{(022)}(t_4)}{M_1} \approx \left( M_1\mu \right)^2.
\end{align}
Inserting $M_1=0.9368 M_0$, the mass of the $\{0,2,2\}$ mode at $t_4$ is $8.22\times 10^{-3} M_0$. The value from solving the differential equations is $9.59\times 10^{-3} M_0$.

The timescales can also be estimated. Using Eq.~\eqref{eq:Ms_nlm} and $\omega_I^{nlm} \sim \mu(M_0\mu)^{4l+5}$, the time $t_1$ can be estimated with,
\begin{align}
\frac{t_1}{M_0} \sim \frac{1}{2(M_0\mu)^{4l+6}}\log\frac{M_s^{(011)}(t_1)}{M_{s0}^{(011)}}.
\end{align}
For our chosen parameters, it gives $t_1 \sim 2.04\times 10^{10} M_0$, while the numerical method gives $7.03\times 10^{11} M_0$. For the value of $t_2$, note that the BH spin between $t_1$ and $t_2$ can be estimated with Eq.~\eqref{eq:111_attractor} at time $t_1$. Especially, we look for a solution with $\omega_I^{(011)}>0$ and $\omega_I^{(111)}<0$. Such a solution does not exist with the LO analytic approximation of $\omega_I^{(nlm)}$. Using the NLO approximation in Ref.~\cite{Bao:2022hew}, one obtains $M_0\omega^{011}_I(t_1) = 7.57\times 10^{-17}$. Then the time $t_2$ could be estimated by,
\begin{align}\label{eq:t2}
t_2\approx \frac{M_s^{(111)}(t_1)}{2\,\omega_I^{011}(t_1) M_s^{(011)}(t_1)}.
\end{align}
For our chosen parameters, we get $t_2/M_0 \approx 4.69\times 10^{14}$, which is about $1/3$ of the result $1.45\times 10^{15}$ from the numerical calculation.

It is interesting to ask what is the maximum value of $M_s^{(011)}$ in the evolution of the system. The ratio on the LHS of Eq.~\eqref{eq:ratio_M011_M0} reaches the maximum value of $a_{*0}/16$ at $M_0\mu = a_{*0}/8$. One could then insert this value in Eq.~\eqref{eq:ratio_M111_M0} to get the maximum value of $M_{s}^{(111)}(t_1)$. This procedure can also be applied to modes with $n\geq 2$. Finally, the maximum mass of the $\{0,1,1\}$ mode is the summation of all the $l=m=1$ modes,
\begin{align}
\frac{M_s^{(011)}}{M_0}\lesssim \sum_{n=0}^{\infty} \left(\frac{M_{s0}^{(n11)}}{M_0}\right)^{1-\frac{\beta_{n1}}{\beta_{01}}} \left(\frac{a_{*0}}{16}\right)^{\frac{\beta_{n1}}{\beta_{01}}}.
\end{align}
Either a hard truncation or a mild smearing factor must be given for the initial masses $M_{s0}^{(n11)}$ so that the total condensate mass at $t=0$ is bounded. If choosing $M_{s0}^{(011)} =M_{s0}^{(111)} = 10^{-3}M_0$ and the initial mass of all other modes to be zero, we get $M_s^{(011)}/M_0 \lesssim 0.0668$ at $a_{*0}=1$.

\begin{figure}
	\centering
 \includegraphics[width=0.5\textwidth]{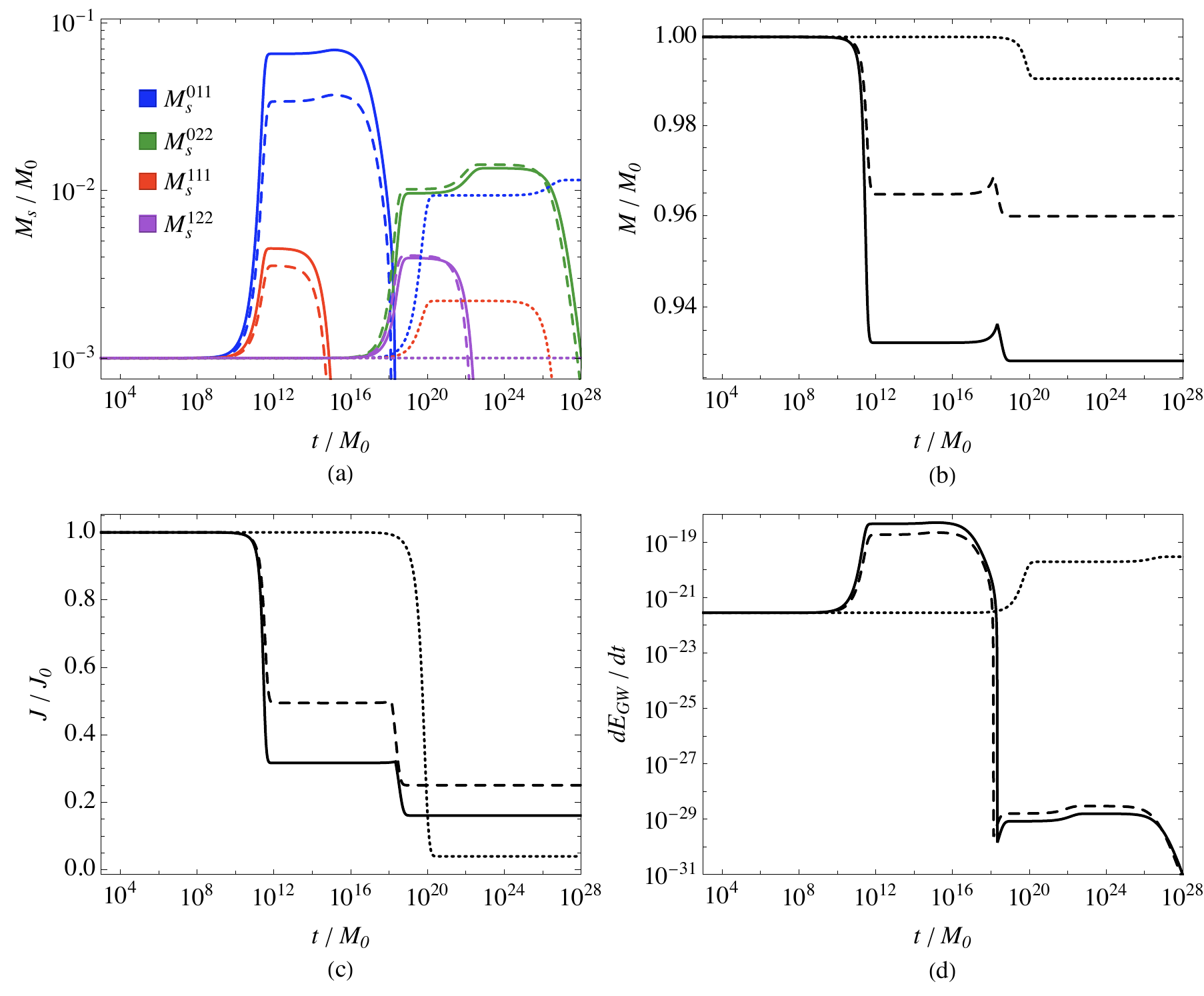}
	\caption{The evolution of the scalar cloud masses, BH mass, BH spin, and GW emission luminosity as a function of time. The solid curves are the baseline case with initial parameters $M_0\mu=0.1$ and $a_{*0}=0.99$, the same as in Fig.~\ref{fig:_evolution-0.1-0.99}. The dashed curves are with initial parameters $M_0\mu=0.1$ and $a_{*0}=0.7$. The dotted curves are with initial parameters $M_0\mu=0.01$ and $a_{*0}=0.9$. The initial mass of each mode is set as $10^{-3}M_0$ for all cases. The effects of all four modes are included for each curve in panels (b), (c), and (d).
	}
	\label{fig:_evolution-diff-ini-para}
\end{figure}

To study the dependence on the initial parameters, we complete two more calculations by varying the BH initial mass or spin. In the first calculation, $M_0\mu$ is fixed at 0.1, while the value of $a_{*0}$ is reduced from $0.99$ to $0.7$. In the second calculation, $a_{*0}$ is fixed at 0.99, while the $M_0\mu$ is reduced from 0.1 to 0.01. The results are shown in Fig.~\ref{fig:_evolution-diff-ini-para}. The curves in Fig.~\ref{fig:_evolution-0.1-0.99} are also plotted as a baseline for comparison. In Tab.~\ref{tab:comparison}, we compare the estimates of the condensate masses and timescales to the values from solving the differential equations. The estimates fit reasonably well for all the quantities.

\begin{table*}
\bgroup
\def\arraystretch{1.5}
\begin{tabular}{c|c c c c c|c c c c c}
\hline
\multirow{2}{1.5em}{Set} &  \multicolumn{5}{c|}{Numerical results} & \multicolumn{5}{c}{Estimates}  \\
 & ${M_s^{(011)}(t_1)}$ & ${M_s^{(111)}(t_1)}$ & ${M_s^{(022)}(t_4)}$ & ${t_1}$ & ${t_2}$ & ${M_s^{(011)}(t_1)}$ & ${M_s^{(111)}(t_1)}$ & ${M_s^{(022)}(t_4)}$ & ${t_1}$ & ${t_2}$ \\
\hline
 1 & $6.51\cdot 10^{-2}$ & $4.50\cdot 10^{-3}$ & $9.59\cdot 10^{-3}$ & $7.03\cdot 10^{11}$ & $1.45\cdot 10^{15}$ & $5.90\cdot 10^{-2}$ & $4.19\cdot 10^{-3}$ & $8.22\cdot 10^{-3}$ & $2.04\cdot 10^{10}$ & $4.69\cdot 10^{14}$ \\
 2 & $3.38\cdot 10^{-2}$ & $3.55\cdot 10^{-3}$ & $1.01\cdot 10^{-2}$ & $9.15\cdot 10^{11}$ & $1.33\cdot 10^{15}$ & $3.00\cdot 10^{-2}$ & $3.30\cdot 10^{-3}$ & $9.03\cdot 10^{-3}$ & $1.70\times 10^{10}$ & $3.35\cdot 10^{14}$ \\
 3 & $9.32\cdot 10^{-3}$ & $2.19\cdot 10^{-3}$ & — & $4.36\cdot 10^{20}$ & $2.71\cdot 10^{27}$ & $9.50\cdot 10^{-3}$ & $2.20\cdot 10^{-3}$ & $9.65\cdot 10^{-5}$ & $1.13\times 10^{20}$ & $3.02\cdot 10^{26}$ \\
\hline
\end{tabular}
\egroup
\caption{Comparison of the estimates of condensate masses and the timescales to the numerical results from solving the differential equations. All values have the unit $M_0$. The initial parameters are $M_0\mu=0.1$ and $a_{*0}=0.99$ for Set 1, $M_0\mu=0.1$ and $a_{*0}=0.7$ for Set 2, and $M_0\mu=0.01$ and $a_{*0}=0.99$ for Set 3. The initial masses of $\{0,1,1\}$, $\{1,1,1\}$, $\{0,2,2\}$ and $\{1,2,2\}$ modes are $10^{-3}M_0$. Other modes are absent in the evolution of the BH-condensate system. The numerical value of $M^{(022)}_s(t_4)$ is not provided because our code is very unstable there.
\label{tab:comparison}
}
\end{table*}

\subsection{With Accretion}\label{subsec:withAcc}

The accretion effect with the dominant modes has been studied in Refs.~\cite{Brito:2014wla, Hui:2022sri}. Below we first interpret the results in Ref.~\cite{Brito:2014wla}, which considers only the $\{0,1,1\}$ mode. In their calculation, the mass accretion is conservatively assumed to be a fraction of the Eddington rate,
\begin{align}
\dot{M}_\text{ACC} \equiv f_\text{Edd} \dot{M}_\text{Edd} \sim 0.02 f_\text{Edd}\frac{M(t)}{10^6 M_\odot} M_\odot \text{yr}^{-1},
\end{align}
where the average value of the radiative efficiency is assumed to be $0.1$. The angular momentum accretion is,
\begin{align}\label{eq:ACC_J_M}
\dot{J}_\text{ACC}\equiv \frac{L(M,J)}{E(M,J)}\dot{M}_\text{ACC},
\end{align}
where $L(M,J)$ and $E(M,J)$ are the angular momentum and energy per unit mass on the innermost stable circular orbit (ISCO) of the Kerr metric, respectively. They are related to the ISCO radius $r_\text{ISCO}(M,J)$ by,
\begin{subequations}
\begin{align}
L(M,J) &= \frac{2M}{3\sqrt{3}}\left(1+2\sqrt{\frac{3r_\text{ISCO}}{M}-2}\right),\\
E(M,J) &= \sqrt{1-\frac{2M}{3r_\text{ISCO}}}.
\end{align}
\end{subequations}
Then, the accretion effect is included by adding $\dot{M}_\text{ACC}$ and $\dot{J}_\text{ACC}$ on the right-hand sides of Eqs.~\eqref{eq:Etot} and \eqref{eq:Jtot}, respectively.

With only the $\{0,1,1\}$ mode, the authors of Ref.~\cite{Brito:2014wla} carefully solved the evolution of the BH and the scalar cloud (see Figs.~2 and 5 in Ref.~\cite{Brito:2014wla}). With the observation that the BH mass grows over time, their numerical result has four distinct phases, which could be well-understood with the dependence of $\omega_I$ on $M$ and $a_*$ (see Fig.~\ref{fig:_anay_Im_omega_a=0.99_to_0.4}). 
\begin{itemize}
\item {\it Accelerating phase}. 
Initially, $M_s^{(011)}$ is small, while the BH mass and spin  $a_{*}$ increase over time because of accretion. The values of $\omega_I^{(011)}$ and $M_s^{(011)}$ increase consequently.
\item {\it Decelerating phase}. 
When $M_s^{(011)}$ is large such that its extraction of the angular momentum overruns the accretion for the BH, the BH loses its spin and $a_{*}$ decreases. The value of $\omega_I^{(011)}$ decreases consequently, while $M_s^{(011)}$ increases at a slower rate.
\item {\it Attractor phase}.
$\omega_I^{(011)}$ decreases to a value such that the extraction of angular momentum nearly balances the accretion for the BH. The value of $a_{*}$ is very close to $a_{*C}^{011}$ due to the large value of $M_s^{(011)}$. In this phase, the BH evolves along the Regge trajectory of the $\{0,1,1\}$ mode to larger mass and spin.
\item {\it Quasi-normal phase}. 
After $a_*$ reaching the maximum value $\sim 0.998$ along the Regge trajectory, moving to larger BH mass results in a negative $\omega_I^{011}$. The mode is then stable. The scalar condensate quickly shrinks and returns its mass and angular momentum to the BH.
\end{itemize}
In the above analysis, we have ignored the GW emission, which would not affect the evolution substantially \cite{Brito:2014wla}. The critical angular momentum $a_{*C}^{nlm}$ could be estimated with Eq.~\eqref{eq:aStarC}. The critical BH mass $M_C^{nlm}$ at which the quasi-normal phase starts could also be estimated by setting $a_{*C}^{nlm} = 1$ in Eq.~\eqref{eq:aStarC}. The accelerating and decelerating phases are separated by the time with $\dot{a}_*=0$. At this time, the value of $\dot{J}$ is still positive. A detailed analysis of these two quantities is given in App.~\ref{app:dot_a_J}.

Before discussing the accretion effect with subdominant modes, we first look at our previous result in Fig.~\ref{fig:_evolution-0.1-0.99}, which could be interpreted as the evolution with a negligible accretion rate. In this case, the early-stage accelerating phase does not exist. The time range $t<t_1$ is the decelerating phase of all modes. The time range $t_1<t<t_2$ is the attractor phase of the $\{1,1,1\}$ mode. As explained before, the BH evolves closer to the $\{0,1,1\}$ Regge trajectory during this time, because of the large mass of the $\{0,1,1\}$ mode. As the $\{1,1,1\}$ mode shrinks, the BH spin reduces and approaches $a_{*C}^{011}$. It is thus the decelerating phase of the other three modes. The time range $t_2<t<t_3$ is the attractor phase of the $\{0,1,1\}$ mode. The BH tightly follows the $\{0,1,1\}$ Regge trajectory. The energy and angular momentum of the $\{0,1,1\}$ mode is mostly dissipated by GWs, while a small part is absorbed by the BH and the $l=m=2$ modes. From energy and angular momentum conservation, the BH mass increases by $\mu$ when two $\{0,1,1\}$ scalars are absorbed by the BH to produce a scalar in the $l=m=2$ modes. The BH spin $a_*$, which is determined by Eq.~\eqref{eq:aStarC}, increases with its mass. It is thus the accelerating phase of the $l=m=2$ modes. Generally, the attractor phases of subdominant modes are the decelerating phases of other modes, while those of dominant modes are the accelerating phases of the modes with larger values of $l$. The later evolution has the same pattern, with $\{1,2,2\}$ and $\{0,2,2\}$ modes entering the attractor phase consecutively. Without accretion, no mode enters the quasi-normal phase.

There are several differences if the accretion is turned on for the multi-mode scenario. In particular, the dominant modes may survive through the attractor phase and be depleted in the quasi-normal phase. We still focus on the scenario with two dominant modes and two subdominant modes using the same initial parameters as in Fig.~\ref{fig:_evolution-0.1-0.99}. In the beginning, there is an accelerating phase for all modes. The scalar condensates grow more rapidly than the curves shown in Fig.~\ref{fig:_evolution-0.1-0.99} at $t<t_1$. The BH mass and spin also increase, with rates depending on the accretion efficiency. Then at a certain time, the angular momentum extraction from the BH overruns the input from accretion, causing the spinning-down of the BH. This is the decelerating phase, in which all modes grow with rates decreasing over time. When the BH spin drops to the value $\sim a_{*C}^{111}$, the $\{1,1,1\}$ mode enters its attractor phase, returning its mass and angular momentum to the BH. The BH follows the $\{1,1,1\}$ attractor which lies between the Regge trajectories of $\{0,1,1\}$ and $\{1,1,1\}$ modes, similar to the case in Fig.~\ref{fig:Attractors}. Nonetheless, with accretion, the BH mass grows from $t_1$ to $t_2$, causing the two Regge trajectories to rise with time as well. If accretion is slow, the BH spin still decreases during this time, and it is the decelerating phase of other modes. If accretion is rapid, however, the BH spin grows and this time range turns out to be the accelerating phase of other modes. After the $\{1,1,1\}$ mode is depleted, the $\{0,1,1\}$ mode enters its attractor phase and the BH follows the $\{0,1,1\}$ Regge trajectory. If the BH mass grows slowly, the $\{0,1,1\}$ mode is depleted in its attractor phase. On the contrary, if the BH mass grows rapidly, this mode could move to the quasi-normal phase before being depleted. Then, in the quasi-normal phase, it is quickly drained. In either case, the $l=m=2$ modes are in the accelerating phase before the $\{0,1,1\}$ scalar is eliminated. Then a similar pattern repeats for the $l=m=2$ modes.

The masses of different modes can also be estimated with the observations above. We still consider the four modes as before and assume that the initial masses of all condensate modes are negligible. In the first accelerating phase, when the masses of all modes are small, the accretion is dominant and the BH mass increases exponentially. Then the system enters the decelerating phase, which is dominated by superradiant instability. The mass of the $l=m=1$ modes at the end of this phase can be estimated in the same way as Eqs.~\eqref{eq:ratio_M011_M0} and \eqref{eq:ratio_M111_M0}. $M_0$ and $a_{*0}$ should be interpreted as the BH mass and spin at the beginning of the decelerating phase, respectively. Following the decelerating phase is the attractor phase of the $\{1,1,1\}$ mode, which is dominated by the $\{0,1,1\}$ instability. The duration of this phase can still be approximated with Eq.~\eqref{eq:t2}. Generally, the $\{1,1,1\}$ attractor phase is not affected much by the accretion as long as the latter is not unrealistically rapid.

Then the system enters the attractor phase of $\{0,1,1\}$. From here on, the accretion makes a big difference. The $l=m=2$ modes are still negligible, so one only needs to consider the $\{0,1,1\}$ mode. Combining the first two equations of Eqs.~\eqref{eq:Evoluton_ODE} with Eq.~\eqref{eq:ACC_J_M} and $J=(a_{*C}^{(011)}+\delta)M^2$, and also using Eq.~\eqref{eq:aStarC} for $a_{*C}^{(011)}$, we arrive at,
\begin{align}\label{eq:accretion_estimate}
\begin{split}
&\left(12M^2\mu^2+2\delta M\mu-\eta\mu\right)\dot{M}+\dot{\delta} M^2\mu \\
&\hspace{2cm}
+(1-\eta \mu)\left(\dot{M}_s^{(011)}+\dot{E}_\text{GW}\right)=0,
\end{split}
\end{align}
where $\eta\equiv L(M,J)/E(M,J)$. Without accretion, all three terms are identically zero. If accretion is present, the evolution is determined by the competition of the accretion and the GW emission. Here we are more interested in the upper limit of $M_s^{(011)}$. For this purpose, we turn off the GW emission in Eq.~\eqref{eq:accretion_estimate} and take its leading order terms. After some algebra, one obtains,
\begin{align}
M_{s,f}^{(011)} - M_{s,i}^{(011)} = \int_{M_i}^{M_f} dM \frac{12M^2\mu^2 -\eta \mu}{\eta\mu-1}.
\end{align}
where the subscripts $i$ and $f$ indicate the quantities at the beginning and the end of the $\{0,1,1\}$ attractor phase, respectively. To estimate $\eta$, we set $r_\text{ISCO} = 6M$, the value for a Schwarzschild BH. The obtained $\eta$ is $3\sqrt{6}M/2$. We further restrict ourselves to $\mu M_i\ll 1$, in which case $\mu M_{s,i}^{(011)}$ can be ignored as well. The attractor phase ends at $a_{*C}\sim 1$, which gives $\mu M_f\sim 0.5$ from Eq.~\eqref{eq:aStarC}. Finally, one could complete the integral and get,
\begin{equation}
{M^{(011)}_{s,f}}/{M_f} \sim 0.517.
\end{equation}
Numerically solving the differential equations gives $0.34$ \cite{Brito:2014wla}. A more accurate value can be achieved using the $r_\text{ISCO}$ in the Kerr spacetime. If the GW emission is turned on, the mass ratio drops to $0.185$ \cite{Brito:2014wla}, indicating the importance of the GW emission in the attractor phase of the $\{0,1,1\}$ mode.

\subsection{Nonlinear Effects and Interference }\label{subsec:nonlinear}

Finally, we discuss some mechanisms not included in the above calculation. The back reaction of the condensate to the metric is negligible in Ref.~\cite{Brito:2014wla}, due to the small energy density of the scalar field. The self-interaction of the scalars changes the shape of the wavefunctions, even causing a bosenova before a mode reaches its maximum mass. Numerical simulation shows that the bosenova collapse happens when the scalar cloud to BH mass ratio $M_s/M$ is approximately $0.16$ \cite{Yoshino:2012kn}. From the calculations above, this ratio is always less than $0.0668$ without accretion but could reach as much as $0.34$ if accretion is present. The study of the BH-condensate evolution with a bosenova is beyond the scope of this work.

Another consequence of the scalar self-interaction is level-mixing, which may shut down some superradiant modes. We focus on the effect of the dominant $\{0,1,1\}$ mode on the subdominant $\{1,1,1\}$ mode. The self-interaction could annihilate two $\{1,1,1\}$ scalars into a scalar in the $\{0,1,1\}$ mode and a scalar with energy $\omega'=2\omega_R^{111}-\omega_R^{011}$ and $m'=1$. With the approximation of $\omega_R$ in Eq.~\eqref{eq:omega_R}, one gets $\omega'>\mu$, implying it is a continuous mode. Following the same argument as in Ref.~\cite{Arvanitaki:2010sy}, the flux at the horizon is,
\begin{align}\label{eq:flux}
\omega_R^{011}\left(\omega_R^{011} - \Omega_H \right) \left|u_h\right|^2
+\omega'\left(\omega' - \Omega_H \right) \left|v_h\right|^2,
\end{align}
where $u_h$ and $v_h$ are the wavefunctions of $\{0,1,1\}$ and the continuous modes at the horizon due to the self-interaction, respectively. We have assumed $\omega_R^{111} = \Omega_H$, when the $\{1,1,1\}$ mode reaches its maximum. In this case, the first term in Eq.~\eqref{eq:flux} is negative, while the second term is positive. The overall sign depends on the values of $u_h$ and $v_h$. Without environmental scalars, the continuous mode wave function is suppressed by $\alpha^2$ compared to the $\{0,1,1\}$ wave function, where $\alpha$ is the coupling of the scalar self-interaction. Thus, we conclude that the level-mixing effect does not terminate the superradiant instability of the $\{1,1,1\}$ mode earlier than $t_1$ in Fig.~\ref{fig:_evolution-0.1-0.99}.

Another omitted effect is the interference in the GW emission. In Eqs.~\eqref{eq:Evoluton_ODE}, only the GW emitted by every single mode is included. GWs can also be produced by the transition from one mode to another, or by the annihilation of two scalars in different modes. The former can be calculated via the quadrupole formula and is negligible \cite{Arvanitaki:2010sy}. The latter process has also been studied for modes with different $l$, such as $\{0,1,1\}$ and $\{0,2,2\}$ modes. From the scaling of the expressions in Eqs.~\eqref{eq:GW-rad}, this interference is suppressed by $(M\mu)^2$ compared to $\dot{E}_\text{GW}^{(011)}$ and is not important. To the contrary, the interference of the $\{0,1,1\}$ and $\{1,1,1\}$ modes is not suppressed. The signal is the strongest between $t_1$ and $t_2$ in Fig.~\ref{fig:_evolution-0.1-0.99}, causing a beat feature of the GW waveform, which is the topic of the next section. 
\\

\section{Effects on GW Emission}
\label{sec:GW_emission}

In this section, we study the effects of the subdominant modes on GW emission. In particular, we focus on the interference between the $\{0,1,1\}$ and $\{1,1,1\}$ modes. Due to the small difference in the periods,  constructive interference happens when the dense regions of these two modes overlap. On the other hand,  destructive interference happens when they are out of phase. This results in a periodic modulation of the GW emission flux. It could be used as a special feature of some BH-condensate systems, to distinguish them from other continuous sources, such as rotating neutron stars. In the context below, we first introduce the calculation framework in Sec.~\ref{subsec:GW_framework}, then calculate the GW emission of a single mode in Sec.~\ref{subsec:GW_single} and compare our results with the approximations in Eqs.~\eqref{eq:GW-rad}. After that, we discuss in detail the beat-like pattern due to the interference of different modes in Sec.~\ref{subsec:_Multiple_modes}.\\

\subsection{Calculation Framework}\label{subsec:GW_framework}

We follow the method in Ref.~\cite{Brito:2017zvb} to calculate the GW emission luminosity based on the Newman-Penrose (NP) formalism in Kerr spacetime \cite{Newman:1961qr}. The complex null tetrads $\{l^\mu, n^\mu, m^\mu, m^{*\mu}\}$ are defined as,
\begin{align}
	l^{\mu} &=\left[\left(r^{2}+a^{2}\right) / \Delta, 1,0, a / \Delta\right],\\
	n^{\mu} &=\left[r^{2}+a^{2},-\Delta, 0, a\right] /(2 \Sigma), \\
	m^{\mu} &=[i a \sin \theta, 0,1, i / \sin \theta] \left/\left[\sqrt 2(r+i a \cos \theta)\right]\right.,
	  \label{eq:null_tetrad}
  \end{align}
and $m^{*\mu}$ is the complex conjugate of $m^\mu$. Their scalar products vanish, except $l^\mu n_\mu=1$ and $m^\mu m^*_\mu=-1$.

In NP formalism, Einstein's equation can be transformed into Teukolsky equations \cite{Teukolsky:1972my,Teukolsky:1973ha,Press:1973zz}. The ten independent components of the Weyl tensor are converted into five complex Weyl scalars. Among them, we are interested in the Weyl scalar $\psi_4$ which is interpreted as the outgoing transverse radiation \cite{Szekeres:1965ux}. We further define $\psi=\rho^{-4}\psi_4$ with $\rho = -\left(r-ia\cos\theta\right)^{-1}$, which satisfies the Teukolsky equation, 
\begin{widetext}
\begin{equation}
\begin{aligned}
&\left[\frac{\left(r^{2}+a^{2}\right)^{2}}{\Delta}-a^{2}\sin^{2}\theta\right]\frac{\partial^{2}\psi}{\partial t^{2}}+\frac{4Mar}{\Delta}\frac{\partial^{2}\psi}{\partial t\partial\varphi}+\left[\frac{a^{2}}{\Delta}-\frac{1}{\sin^{2}\theta}\right]\frac{\partial^{2}\psi}{\partial\varphi^{2}}-\Delta^{2}\frac{\partial}{\partial r}\left(\Delta^{-1}\frac{\partial\psi}{\partial r}\right)
-\frac{1}{\sin\theta}\frac{\partial}{\partial\theta}\left(\sin\theta\frac{\partial\psi}{\partial\theta}\right)\\
&\hspace{2cm}
+4\left[\frac{a(r-M)}{\Delta}+\frac{i\cos\theta}{\sin^{2}\theta}\right]\frac{\partial\psi}{\partial\varphi}+4\left[\frac{M\left(r^{2}-a^{2}\right)}{\Delta}-r-ia\cos\theta\right]\frac{\partial\psi}{\partial t}
+\left(4\cot^{2}\theta+2\right)\psi =4\pi\Sigma T,
\end{aligned}
\label{eq:Teukolsky_eq}
\end{equation} 
where $T=2\rho^{-4}T_4$, and $T_4$ is defined as \cite{Sasaki:2003xr},
\begin{equation}
	\begin{aligned}
		T_4=&-\frac{1}{2} \rho^{8} \rho^* J_{\theta,-1}\left[\rho^{-4} J_{\theta,0}\left(\rho^{-2} \rho^{*-1} T_{n n}\right)\right] +\frac{1}{2 \sqrt{2}} \rho^{8} {\rho}^* \Delta^{2} J_{\theta,-1}\left[\rho^{-4} {\rho}^{*2} J_{r}\left(\rho^{-2} {\rho}^{*-2} \Delta^{-1} T_{{m}^* n}\right)\right] \\
		&-\frac{1}{4} \rho^{8} {\rho}^* \Delta^{2} J_{r}\left[\rho^{-4} J_{r}\left(\rho^{-2} {\rho}^* T_{{m^* m^*}}\right)\right] +\frac{1}{2 \sqrt{2}} \rho^{8} {\rho}^* \Delta^{2} J_{r}\left[\rho^{-4} {\rho}^{*2} \Delta^{-1} J_{\theta,-1}\left(\rho^{-2} {\rho}^{^*-2} T_{{m}^* n}\right)\right],
	\end{aligned}
\label{eq:def_T4}
\end{equation}
\end{widetext}
where, 
\begin{subequations}\label{eq:Js}
\begin{align}
	J_{\theta,k}=&\partial_{\theta}-\frac{i}{\sin\theta}\partial_\varphi - i a\sin\theta\partial_t +k\cot\theta,\\
	J_r =& \partial_{r}- \frac{1}{\Delta}\left[(r^2+a^2)\partial_t+a\partial_\varphi\right],
\end{align}
\end{subequations}
and the tetrad components of the stress-energy tensor are $T_{ab} \equiv T_{\mu\nu} a^\mu b^\nu$ with $a,b=\{l,n,m,m^*\}$.

The variables of $\psi$ can be separated in the form of,
\begin{align}
	\psi & =\int d\widetilde\omega\sum_{\widetilde{l},\widetilde{m}}R_{\widetilde{l} \widetilde{m}}(r){}_{-2}S_{\widetilde{l} \widetilde{m}}(\theta)\dfrac{e^{i \widetilde{m} \varphi}}{\sqrt{2\pi}}e^{-i\widetilde\omega t}.
\label{eq:separate_psi}		
\end{align}
By inserting this into Eqs.~\eqref{eq:Js}, $\partial_\varphi$ and $\partial_t$ can be replaced by $i\widetilde{m}$ and $-i\widetilde{\omega}$, respectively. Then, $J_{\theta,k}$ only depends on $\theta$ while $J_r$ only depends on $r$, which is indicated by the subscripts. The function $_{-2}S_{\widetilde{l}\widetilde{m}}(\theta)$ is the eigenfunction of the angular part with eigenvalue ${}_{-2}A_{\widetilde{l} \widetilde{m}}$, satisfying,
\begin{equation}
	\begin{aligned} 
	&\frac{1}{\sin\theta}\frac{d}{d\theta}\left[\sin\theta\frac{d_{-2}S_{\widetilde{l} \widetilde{m}}(\theta)}{d\theta}\right]
	+\left[(a\widetilde\omega\cos\theta+2)^2-\frac{\widetilde{m}^{2}}{\sin^{2}\theta}\right.\\
	 &\hspace{1cm}
	 \left.+\frac{4\widetilde m\cos\theta}{\sin^{2}\theta}-4\cot^{2}\theta-6+{}_{-2}A_{\widetilde{l} \widetilde{m}}\right]{}_{-2}S_{\widetilde{l}\widetilde{m}}(\theta)
	 =0.
	\end{aligned}
	\label{eq:spin-weighted_S}
\end{equation}
The orthonormal condition is \cite{Berti:2005gp},
\begin{equation}
	\int_{-1}^{1} d \cos\theta\, {}_{-2} S_{\widetilde{l} \widetilde{m}}(\theta)\, {}_{-2}S_{\widetilde{l}' \widetilde{m}}(\theta)  = \delta_{\widetilde{l} \widetilde{l}'}.
\end{equation}
In our calculation, the function form of ${}_{-2} S_{\widetilde{l} \widetilde{m}}(x)$ is calculated using the Black Hole Perturbation Toolkit \cite{BHPToolkit:Manual} based on Leaver's continued fraction method \cite{Leaver:1985ax}. We will come back to the equation for ${R}_{\widetilde{l}\widetilde{m}}$ later.

For our purpose, the frequency distribution $f(\omega)$ in Eq.~\eqref{eq:_wave_function_1} has discrete values,
\begin{align}
f(\omega)=\sum_{nlm}\sqrt{N_{nlm}}\delta(\omega-\omega^{nlm}).
\end{align}
Hereinafter, we ignore the small imaginary part of $\omega^{nlm}$. The stress-energy tensor of the scalar field $\Phi$ defined in Eq.~\eqref{eq:_wave_function_1} is, 
\begin{equation}\label{eq:Tmunu}
	\begin{aligned}
		T_{\mu\nu}=\sum_{i,j} \dfrac{1}{2}\sqrt{\dfrac{N_{i}N_{j}}{\omega_{i}\omega_{j}}}&\bigg[\mathcal{T}_{\mu\nu}\left(\phi_i,\phi_j\right)+\mathcal{T}_{\mu\nu}\left(\phi^*_i,\phi_j\right)\\
		&\quad +\mathcal{T}_{\mu\nu}\left(\phi_i,\phi^*_j\right)+\mathcal{T}_{\mu\nu}\left(\phi^*_i,\phi^*_j\right)\bigg],
	\end{aligned}
\end{equation}
where, for compactness, we use $i$ and $j$ to label the condensate modes $\{n,l,m\}$. If $i=j$, it is the contribution from a single mode. The interference of two modes is included in the cross terms with $i\neq j$. The function $\mathcal{T}_{\mu\nu}\left(\phi_i,\phi_j\right)$ is defined as,
\begin{align}
\begin{split}
\mathcal{T}_{\mu\nu}\left(\phi_i,\phi_j\right) =& \frac{1}{2}\bigg[\partial_{\mu}\phi_i\partial_{\nu}\phi_j -g_{\mu\nu}\left(\frac{1}{2}\partial_{\rho}\phi_i\partial^{\rho}\phi_j-\frac{1}{2}\mu^{2}\phi_i\phi_j\right)\\
&\hspace{0.5cm}
+(i\leftrightarrow j)\bigg],
\end{split}
\end{align}
and the other three $\mathcal{T}$'s in the square bracket of Eq.~\eqref{eq:Tmunu} are obtained by replacing the corresponding $\phi$ with its complex conjugate. By inserting Eq.~\eqref{eq:Tmunu} into Eq.~\eqref{eq:def_T4}, the $T_4$ can also be written as the addition of four terms,
\begin{equation}\label{eq:T4}
	\begin{aligned}
		T_{4}=\sum_{i,j} \dfrac{1}{2}\sqrt{\dfrac{N_{i}N_{j}}{\omega_{i}\omega_{j}}} & \bigg[\mathcal{T}_{4}\left(\phi_i,\phi_j\right)+\mathcal{T}_{4}\left(\phi^*_i,\phi_j\right)\\
		&\quad+\mathcal{T}_{4}\left(\phi_i,\phi^*_j\right)+\mathcal{T}_{4}\left(\phi^*_i,\phi^*_j\right)\bigg],
	\end{aligned}
\end{equation}
where $\mathcal{T}_4$ is defined in the same way as Eq.~\eqref{eq:def_T4}, but with $T_{\mu\nu}$ replaced by the corresponding $\mathcal{T}_{\mu\nu}$. The three tetrad components of $\mathcal{T}_{\mu\nu}$ needed in our calculation are given in App.~\ref{app:Tmunu}. 

Now, we discuss the radial function $R_{\widetilde{l}\widetilde{m}}$ defined in Eq.~\eqref{eq:separate_psi}. In our case, the source is the addition of different superradiance modes, each with a specific frequency $\omega$. As a result, the functions $\psi$ and $\psi_4$ also have discrete frequencies. For any two distinct superradiant modes $i$ and $j$, they contribute to $\psi$ and $\psi_4$ four frequencies, $\pm(\omega_i+\omega_j)$ and $\pm(\omega_i-\omega_j)$, from the four terms in Eq.~\eqref{eq:T4}. For a single mode $i$, it contributes by itself only two frequencies, $\pm 2\omega_i$. We first look at the component with $\widetilde{\omega} = \omega_i +\omega_j$ in $\psi$, which also requires $\widetilde{m} = m_i+m_j$. The radial Teukolsky equation for this $R_{\widetilde{l}\widetilde{m}}$ is,
\begin{align}
\begin{split}
	&\Delta^{2}\frac{\partial}{\partial r}\Big(\frac{1}{\Delta}\frac{\partial R^{(i+j)}_{\widetilde{l}\widetilde{m}}(r)}{\partial r}\Big)+\Big[\ensuremath{\dfrac{\widetilde{K}^{2}+4i(r-M)\widetilde{K}}{\Delta}}	\\
&\hspace{1cm}
-8i\widetilde\omega r
 -{}_{-2}\lambda_{\widetilde{l}\widetilde{m}}
 \Big]R^{(i+j)}_{\widetilde{l}\widetilde{m}}(r)
=-G^{(i+j)}_{\widetilde{l}\widetilde{m}}(r),
\end{split}
	\label{eq:_radial_Teq_s=-2}
\end{align}
with,
\begin{subequations}
  \begin{align}
  \widetilde{K} &= \left(r^{2}+a^{2}\right) \widetilde\omega-\widetilde{m} a,\\
  {}_{-2}\lambda_{\widetilde{l}\widetilde{m}} &= {}_{-2}A_{\widetilde{l}\widetilde{m}}+a^2 \widetilde{\omega}^2-2 a \widetilde{m} \widetilde{\omega},\\
	G^{(i+j)}_{\widetilde{l}\widetilde{m}}(r) &= \frac{4}{\sqrt{2\pi}}\!\!\int\!\! dt d\Omega\frac{\Sigma}{\rho^4}\mathcal{T}_{4}(\phi_i,\phi_j)  
	{}_{-2}S_{\widetilde{l}\widetilde{m}}(\theta)e^{i(\widetilde\omega t-\widetilde m\varphi)}.
  \end{align}
\end{subequations}
The superscript $(i+j)$ reminds us that it is the component with frequency $\widetilde{\omega}=\omega_i+\omega_j$.

\eref{eq:_radial_Teq_s=-2} can be solved with Green's function method \cite{Sasaki:2003xr,Brito:2017zvb}. One of the two Green's functions satisfies the boundary condition,
\begin{equation}
	g_{\widetilde{l}\widetilde{m}}^{(i+j)}\rightarrow
	\begin{cases}
	\Delta^{2}e^{-i\widetilde{k} r_{*}} & \text{ for }r\rightarrow r_{+},\\
	r^{3}B^{\mathrm{out}}_{\widetilde{l}\widetilde{m}}e^{i\widetilde\omega r_{*}}+r^{-1}B^{\mathrm{in}}_{\widetilde{l}\widetilde{m}}e^{-i\widetilde\omega r_{*}} & \text{ for }r\rightarrow+\infty.
	\end{cases}
\end{equation}
where $\widetilde{k} = \widetilde{\omega}-\widetilde{m}\Omega_H$ and the coefficients $B^{\mathrm{in}}_{\widetilde{l}\widetilde{m}}$ and $B^{\mathrm{out}}_{\widetilde{l}\widetilde{m}}$ are determined by solving the differential equation from the outer horizon to infinity. The tortoise coordinate $r_*$ is defined as, 
\begin{equation}\label{eq:rStar}
  r_*=r-\frac{r_{+}+r_{-}}{r_{+}-r_{-}}\bigg[r_{-} \log \left(\frac{r-r_{-}}{2M}\right)-r_{+} \log \left(\frac{r-r_{+}}{2M}\right)\bigg].
\end{equation}
Solving for this Green's function is nontrivial. We leave the details in App.~\ref{sec:_Solving_RH_and_Bin}. After all these steps, one could obtain $R^{(i+j)}_{\widetilde{l}\widetilde{m}}$ at infinity,
\begin{equation}\label{eq:R_asym_i+j}
	\lim_{r\rightarrow\infty}R^{(i+j)}_{\widetilde{l}\widetilde{m}}=
U_{\widetilde{l}\widetilde{m}}^{(i+j)} r^{3}e^{i\widetilde\omega r_{*}},
\end{equation}
where the coefficient is,
\begin{align}
U_{\widetilde{l}\widetilde{m}}^{(i+j)} = \frac{i}{2\widetilde\omega B^{\mathrm{in}}_{\widetilde{l}\widetilde{m}}}\int_{r_{+}}^{\infty}dr^{\prime}\frac{g_{\widetilde{l}\widetilde{m}}^{(i+j)}(r^\prime)}{\Delta^{2}}G^{(i+j)}_{\widetilde{l}\widetilde{m}}(r^\prime).
\end{align}
The other three contributions, labelled by the superscripts $-i+j$, $i-j$, and $-i-j$, can be obtained in the same way. For $i=j$, only $2i$ and $-2i$ need to be calculated.  

Each pair of the superradiant modes $i$ and $j$ contributes four addends ($i\neq j$) or two addends ($i=j$) in $\psi_4$. After calculating the contribution from each pair of superradiant modes, we finally obtain $\psi_4$ at infinity by summing all these addends,
\begin{align}
\begin{split}
\lim_{r\rightarrow\infty}\psi_4 &= 
\frac{1}{2\sqrt{2\pi}r}
\sum_{i,j} \sqrt{\frac{N_{i}N_{j}}{\omega_{i}\omega_{j}}} \\
&\hspace{1cm}
\times\sum_{\widetilde{l},\widetilde{m},\widetilde{\omega}}
U_{\widetilde{l}\widetilde{m}}^{(\widetilde{\omega})} {}_{-2}S_{\widetilde{l} \widetilde{m}}(\theta) e^{i\widetilde\omega (r_{*}-t)}e^{i \widetilde{m} \varphi},
\end{split}
\end{align}
where the sums of $\widetilde{m}$ and $\widetilde{\omega}$ run over only those values which can be obtained from the pair $(i,j)$.

From the relation,
\begin{align}
  \lim_{r\rightarrow\infty}\psi_4  = \left(\ddot{h}_{+}-i \ddot{h}_{\times}\right)/2,
\end{align}
one could then obtain the two GW strains at infinity,
\begin{align}
\begin{split}
	\lim_{r\rightarrow \infty}h_{+} &=-\frac{1}{\sqrt{2\pi}r}
\sum_{i,j} \sqrt{\frac{N_{i}N_{j}}{\omega_{i}\omega_{j}}}
	\sum_{\widetilde{l}, \widetilde{m},\widetilde{\omega}} \frac{\left|U_{\widetilde{l}\widetilde{m}}^{(\widetilde{\omega})}\right|}{\widetilde\omega^2}{}_{-2}S_{\widetilde{l} \widetilde{m}}(\theta)\\
&\hspace{1.8cm}
\times\cos\left[\widetilde\omega\left(t-r_*\right)-\widetilde{m}\varphi-\widetilde\phi_{\widetilde{l}\widetilde{m}}^{(\widetilde{\omega})}\right],
\end{split}
\end{align}
where $\widetilde{\phi}^{(\widetilde{\omega})}_{\widetilde{l} \widetilde{m}}=\arg \left(U^{(\widetilde{\omega})}_{\widetilde{l} \widetilde{m}}\right)$. The expression for $h_\times$ is the same, only with the cosine replaced by the sine. Finally, one obtains the GW emission luminosity,
\begin{align}\label{eq:GW_Lumi}
\begin{split}
\frac{dE_\text{GW}}{dt}	&= \frac{r^2}{16\pi}\int d\Omega \,\Big\langle\dot{h}_+^2+\dot{h}_\times^2\Big\rangle\\ 
&= \dfrac{1}{16\pi}
\sum_{i,j,i',j'}\sqrt{\frac{N_i N_j N_i' N_j'}{\omega_i \omega_j \omega_i' \omega_j'}}
\sum_{\widetilde{l}\widetilde{m}\widetilde{\omega}\widetilde{\omega}'}
\frac{\left|U_{\widetilde{l}\widetilde{m}}^{(\widetilde{\omega})}\right| 
\left|U_{\widetilde{l}\widetilde{m}}^{(\widetilde{\omega}')*}\right|}{\widetilde{\omega}\widetilde{\omega}'}\\
&\hspace{0.5cm}
\times\left<\cos\left[\left(\widetilde{\omega}-\widetilde{\omega}'\right)(t-r)-\phi_{\widetilde{l}\widetilde{m}}^{(\widetilde{\omega})}+\phi_{\widetilde{l}\widetilde{m}}^{(\widetilde{\omega}')}\right]\right>,
\end{split}
\end{align}
where $\left< ... \right>$ denotes an average over several GW wavelengths, which constrains $|\widetilde{\omega}-\widetilde{\omega}'|\ll \widetilde{\omega}$ in the summation of the frequencies. In the special case in which one superradiant mode dominates, the GW is monochromatic with frequency $\widetilde\omega$. Then the emission energy flux is reduced to a more familiar form \cite{Teukolsky:1973ha},
\begin{equation}
	\frac{d^{2}E_{\text{GW}}}{dtd\Omega}=\lim_{r\rightarrow\infty}\frac{r^{2}}{4\pi\widetilde\omega^{2}}\left|\psi_{4}\right|^{2}.
\label{eq:power}
\end{equation}
In general, the interference between different modes cannot be simply dropped.

\subsection{GW Emission with a Single Mode}\label{subsec:GW_single}

In this subsection, we consider a single superradiant mode $\{n_0,l_0,m_0\}$ with eigenfrequency $\omega_0$. The total number of scalars in the cloud is $N_0$, so the $f(\omega)$ in Eq.~\eqref{eq:_wave_function_1} is $\sqrt{N_0}\delta(\omega-\omega_0)$. This mode contributes two identical addends with frequencies $\widetilde{\omega} = \pm2\omega_0$ on the right-hand side of Eq.~\eqref{eq:GW_Lumi}. Then the GW emission luminosity is,
\begin{align}\label{eq:GW_Lumi_Single}
\left(\frac{M}{M_{s}}\right)^{2}\frac{dE_\text{GW}}{dt} = \frac{M^2}{32\pi \omega_0^6} \sum_{\widetilde{l}} \left| U_{\widetilde{l},2m_0}^{(2\omega_0)}\right|^2,
\end{align}
where $M_s=N_0 \omega_0$ is the total mass of the condensate.

\begin{figure}
	\includegraphics[width=0.45\textwidth]{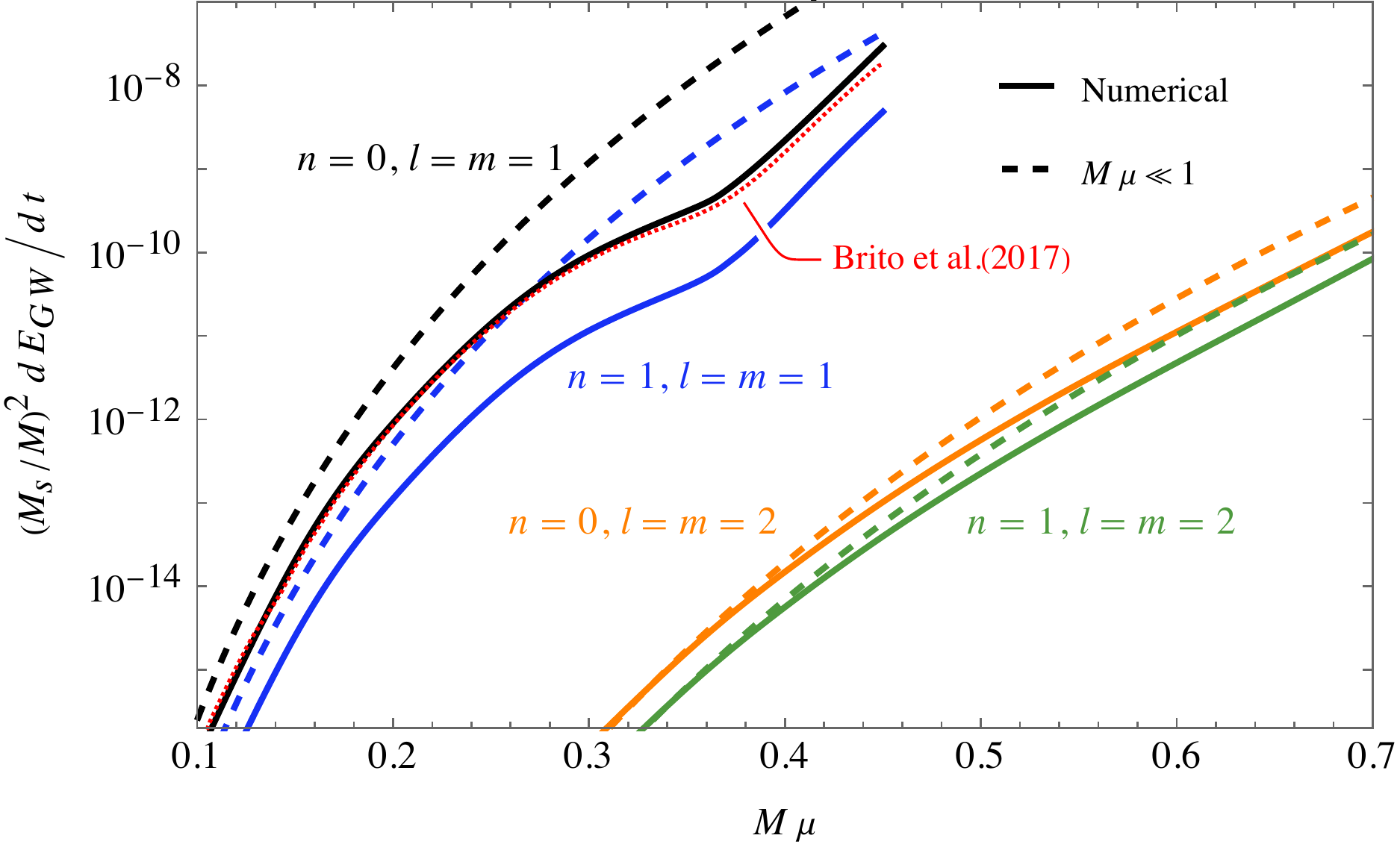}
	\caption{
The GW emission luminosity as a function of the multiplication of the BH mass $M$ and the scalar particle mass $\mu$. Four superradiant modes $\{0,1,1\}$ (black), $\{1,1,1\}$ (blue), $\{0,2,2\}$ (orange), and $\{1,2,2\}$ (green) are calculated in isolated fashion. The interference between different modes is not considered. For each mode, the numerical result with the first four nonzero partial waves (solid curves) is compared to the result in Eqs.~\eqref{eq:GW-rad} (dashed curves). Also shown is the numerical result with $\widetilde{l}=2,3$ for the $\{0,1,1\}$ mode from Ref.~\cite{Brito:2017zvb}. The BH angular momentum $a_{*0}$ is fixed to be its critical value in Eq.~\eqref{eq:aStarC} for each mode.
}
	\label{fig:_dE/dt_vs_brito}
\end{figure}

The frequency $\omega_0$ could be estimated with Eq.~\eqref{eq:omega_R} in the $M\mu\ll 1$ limit. Then the right-hand side is a function of $M\mu$ and the angular momentum $a_*$. With this approximation, we plot the GW emission luminosity as a function of $M\mu$ in Fig.~\ref{fig:_dE/dt_vs_brito}. The angular momentum $a_*$ for each curve is chosen to be the corresponding $a_{*C}^{nlm}$ given in Eq.~\eqref{eq:aStarC}. In the figure, numerical results with the first four nonzero partial waves are compared to the asymptotic expressions given in Eqs.~\eqref{eq:GW-rad}. They agree at $M\mu\ll 1$, showing the consistency of our calculation. At $M\mu=0.1$, the asymptotic expression for the $\{0,1,1\}$ mode is $3$ times larger than the numerical result. Away from the region $M\mu \ll 1$, the asymptotic expression could overestimate the emission power by a factor of as much as $40$ for $l=m=1$ modes and $2.5$ for $l=m=2$ modes. Also plotted in the figure is the numerical solution from Ref.~\cite{Brito:2017zvb}, which keeps only the first two terms in the $\widetilde{l}$ summation for the $\{0,1,1\}$ mode. It differs from our result by at most $39.0\%$, exhibiting the good convergence of the $\widetilde{l}$ summation.

If we were using the more accurate numerical results of the GW emission in the BH evolution in Sec.~\ref{sec:BH_evo}, the overall picture would not change. The evolution process at $t<t_2$ should be the same qualitatively, since the GW emission is negligible in this time range. The GW emission is important in the $\{0,1,1\}$ attractor phase between $t_2$ and $t_3$, when the $\{0,1,1\}$ mode is dissipated mainly by GW emission. Reducing the GW emission by a factor of $3$  in the calculation would extend $t_3-t_2$ by roughly the same factor. Then the $t_3$ calculated in Sec.~\ref{sec:BH_evo} can be taken as a lower limit of the realistic case.  

\subsection{GW Emission with Interference}
\label{subsec:_Multiple_modes}

By solving the BH-condensate evolution with $M_0\mu=0.1$ and $a_{*0}=0.99$, we find that the $\{1,1,1\}$ mode has a mass as much as $6.91\%$ of the $\{0,1,1\}$ mode. In addition, this mass ratio does not change much for a very long time (see Fig.~\ref{fig:_evolution-0.1-0.99}). The coexistence of these two modes gives rise to a modulation of the GW emission. The period of this modulation in the source frame can be estimated using Eq.~\eqref{eq:omega_R},
\begin{align}\label{eq:T_mod}
\begin{split}
T_{\text{mod}} &= \frac{2\pi}{\omega^{(111)} - \omega^{(011)}}
=2880  \left(\frac{0.1}{M\mu}\right)^2 T_\text{GW}\\
&\approx 6.0\times 10^4 \,\text{sec} \left(\frac{10^{-16}\text{eV}}{\mu}\right) \left(\frac{0.1}{M\mu}\right)^2,
\end{split}
\end{align}
where $T_\text{GW}=\pi/\mu$ is approximately the GW period in the source frame. In the particle picture, the strongest GW component is from the annihilation of two $\{0,1,1\}$ scalars to a graviton with frequency $2\omega^{(011)}$. The amplitude is proportional to $N_{011}$, where $N_{nlm}$ is the total number of scalars in the $\{n,l,m\}$ mode. The second strongest GW component is from the annihilation of a $\{0,1,1\}$ scalar and a $\{1,1,1\}$ scalar to a graviton with frequency $\omega^{(011)}+\omega^{(111)}$. The amplitude is proportional to $\sqrt{N_{011}N_{111}}$. Since the GW emission luminosity depends quadratically on the amplitude, the strongest interference term is proportional to $N_{011}^{3/2} N_{111}^{1/2}$.

Next, we study this interference quantitatively. The distribution $f(\omega)$ in Eq.~\eqref{eq:_wave_function_1} is,
\begin{align}
f\left(\omega\right)=\sqrt{N_{011}}\delta(\omega-\omega^{(011)})+\sqrt{N_{111}}\delta(\omega-\omega^{(111)}),
\end{align}
where the small imaginary parts of the frequencies are ignored. There are eight frequencies $\widetilde{\omega}$ contributing to Eq.~\eqref{eq:GW_Lumi}. For convenience, we define $\widetilde{\omega}_1 \equiv 2\omega^{(011)}$, $\widetilde{\omega}_2\equiv 2\omega^{(111)}$, $\widetilde{\omega}_3 \equiv \omega^{(011)}+\omega^{(111)}$, and $\widetilde{\omega}_4 \equiv \omega^{(111)}-\omega^{(011)}$. Then the eight frequencies are $\pm\widetilde{\omega}_1$, $\pm\widetilde{\omega}_2$, $\pm\widetilde{\omega}_3$, and $\pm\widetilde{\omega}_4$. From Eq.~\eqref{eq:omega_R}, there is a relation in the $M\mu\ll 1$ limit,
\begin{align}
\omega^{(111)} - \omega^{(011)} \ll \omega^{(011)} \approx \omega^{(111)}.
\end{align}
Three interference terms survive after the averaging in Eq.~\eqref{eq:GW_Lumi},
\begin{widetext} 
\begin{equation}
\begin{aligned}
\frac{dE_\text{GW}}{dt}
=&\frac{1}{8\pi}
{\sum_{\widetilde{l}}}
\Bigg\{
\dfrac{N_{011}^{2}}{{\omega^{(011)}}^2} \frac{\left|U_{\widetilde{l} 2}^{(\widetilde{\omega}_1)}\right|^{2}}{\widetilde{\omega}_{1}^2}
+\dfrac{N_{111}^{2}}{{\omega^{(111)}}^2}\frac{\left|U_{\widetilde{l} 2}^{(\widetilde{\omega}_2)}\right|^{2}}{\widetilde{\omega}_{2}^2} 
+4\dfrac{N_{011}N_{111}}{\omega^{(011)}\omega^{(111)}} \frac{\left|U_{\widetilde{l} 2}^{(\widetilde{\omega}_3)}\right|^{2}}{\widetilde{\omega}_{3}^2}\\
& +4 \sqrt{\dfrac{N_{011}^3N_{111}}{{\omega^{(011)}}^3\omega^{(111)}}} 
\frac{\left|U_{\widetilde{l}2}^{(\widetilde{\omega}_1)}\right| \left|U_{\widetilde{l}2}^{(\widetilde{\omega}_3)}\right|}{\widetilde{\omega}_{1} \widetilde{\omega}_{3}}\cdot \cos\left[\widetilde{\omega}_4\left(t-r_*\right)
-\phi_{\widetilde{l}2}^{(\widetilde{\omega}_3)}+\phi_{\widetilde{l}2}^{(\widetilde{\omega}_1)}
\right] \\
& +2 \dfrac{N_{011}N_{111}}{\omega^{(011)}\omega^{(111)}} 
\frac{\left|U_{\widetilde{l}2}^{(\widetilde{\omega}_1)}\right| \left|U_{\widetilde{l}2}^{(\widetilde{\omega}_2)}\right|}{\widetilde{\omega}_{1} \widetilde{\omega}_{2}}\cdot {\cos\left[2\widetilde{\omega}_4\left(t-r_*\right)
-\phi_{\widetilde{l}2}^{(\widetilde{\omega}_2)}+\phi_{\widetilde{l}2}^{(\widetilde{\omega}_1)}
\right]} \\
& +4 \sqrt{\dfrac{N_{011}N_{111}^3}{{\omega^{(011)}}{\omega^{(111)}}^3}} 
\frac{\left|U_{\widetilde{l}2}^{(\widetilde{\omega}_2)}\right| \left|U_{\widetilde{l}2}^{(\widetilde{\omega}_3)}\right|}{\widetilde{\omega}_{2} \widetilde{\omega}_{3}}\cdot {\cos\left[\widetilde{\omega}_4\left(t-r_*\right)
-\phi_{\widetilde{l}2}^{(\widetilde{\omega}_2)}+\phi_{\widetilde{l}2}^{(\widetilde{\omega}_3)}
\right]} 
\Bigg\}.
  \end{aligned}
  \label{eq:_dE/dt_011_111}
\end{equation}
\end{widetext}

\begin{figure}
	\includegraphics[width=0.45\textwidth]{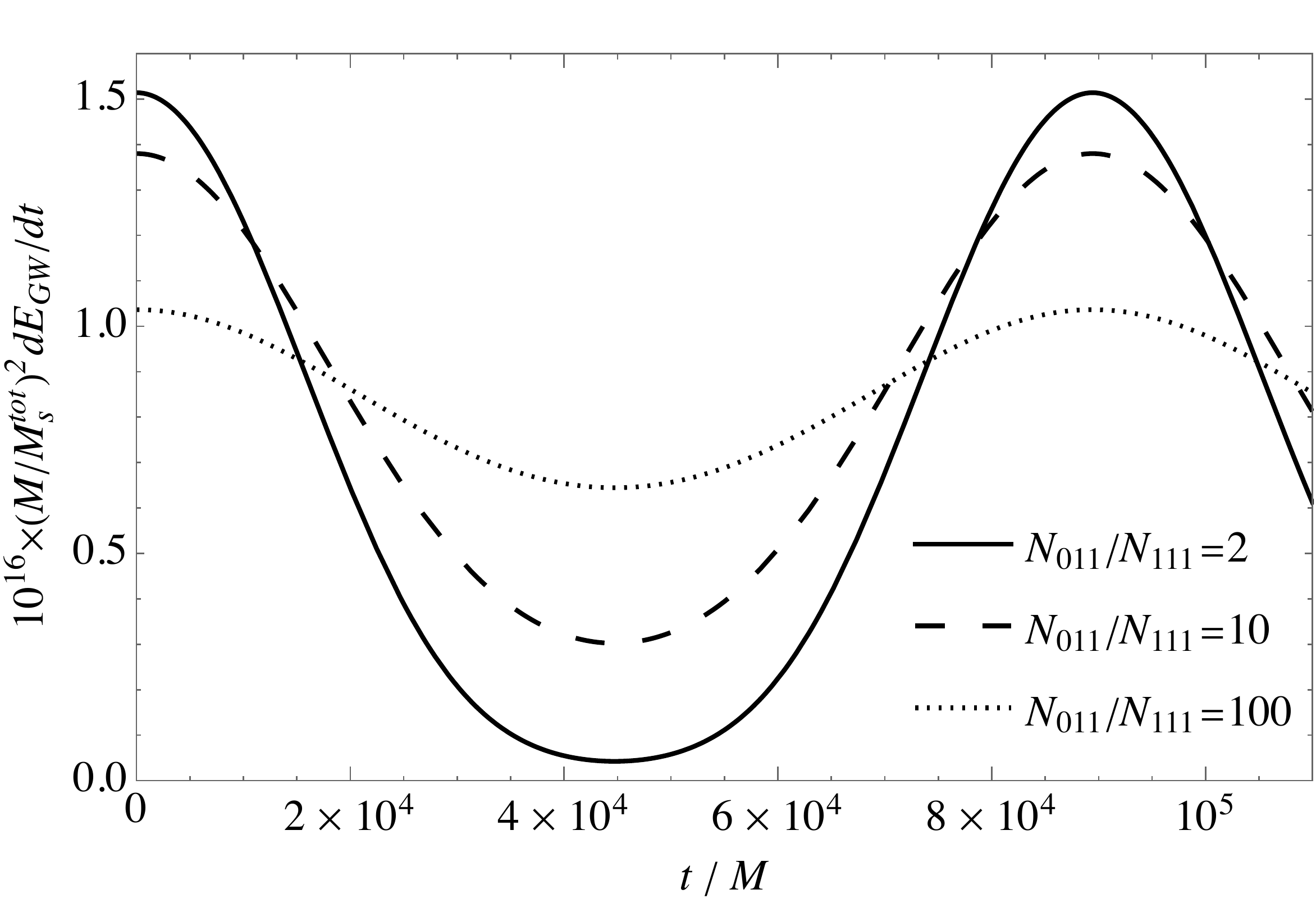}
	\caption{
The beat-like feature of the GW emission luminosity as a function of time, with superradiant modes $\{0,1,1\}$ and $\{1,1,1\}$. The variation of the GW emission luminosity is the interference effect of these two modes. We choose $M\mu=0.1$ and the BH angular momentum $a_{*}$ is fixed to be $a_{*C}^{111}$ in Eq.~\eqref{eq:aStarC}. Three cases are investigated, with $N_{011}/N_{111}$ equal to 2 (solid curve), 10 (dashed), and 100 (dotted). The $M_s^\text{tot}=N_{011}\omega^{(011)} + N_{111}\omega^{(111)}$ is the total mass of the scalar condensates.}
	\label{fig:_GW_beat_emission_energy_flux}
\end{figure}

This GW emission luminosity is plotted as a function of time in Fig.~\ref{fig:_GW_beat_emission_energy_flux}. The parameters are $M\mu=0.1$ and $a_*=0.384$, which is the $a_{*C}^{111}$ calculated using Eq.~\eqref{eq:aStarC}. Three values $2,10,100$ are chosen for $N_{011}/N_{111}$. The two superradiant modes are tuned in phase at $t=0$. We normalize $t$ using the BH mass $M$, which is related to the SI unit by,
\begin{align}
M = 4.93\times 10^{-6} \text{sec}\, \left(\frac{M}{M_\odot}\right).
\end{align}
Instead of the constant GW emission luminosity with a single superradiant mode, here the interference between different modes leads to a beat-like pattern. Not surprisingly, the beat is strong if the masses of the two modes are in the same order. With $N_{011}/N_{111}=2$, the variation of the curve is roughly $94\%$. The curve flattens to a straight horizontal line if the mass of one mode is negligible compared to the other. But the variation is still $\sim 20\%$ when $N_{011}/N_{111}$ is $100$. The behavior could be understood with Eq.~\eqref{eq:_dE/dt_011_111}. The dominant term contributing to a constant GW luminosity scales as $N_{011}^2$. The strongest interference term is only mildly suppressed by $\sqrt{N_{111}/N_{011}}$, explaining the large variation. If we increase the ratio of $N_{111}/N_{011}$, the interference term with frequency $2\widetilde{\omega}_4$ increases in importance, causing a deviation of the curve from a shifted cosine function.

For the parameters used in Fig.~\ref{fig:_evolution-0.1-0.99}, the interference effect changes the GW emission by roughly $10\%$. Including the interference in the evolution of the BH-condensate would not make much difference. Thus, our conclusions in Sec.~\ref{sec:BH_evo} still hold. Nonetheless, the beat-like pattern in the GW emission is a unique feature of those BH-condensate systems in which the subdominant mode has a mass larger than one percent of the dominant mode. It can be used to distinguish some BH-condensate systems from other monochromatic GW sources, supplementary to the frequency drift proposed in Refs.~\cite{Arvanitaki:2014wva,Baryakhtar:2017ngi}.

\begin{figure}[htb]
	\includegraphics[width=0.45\textwidth]{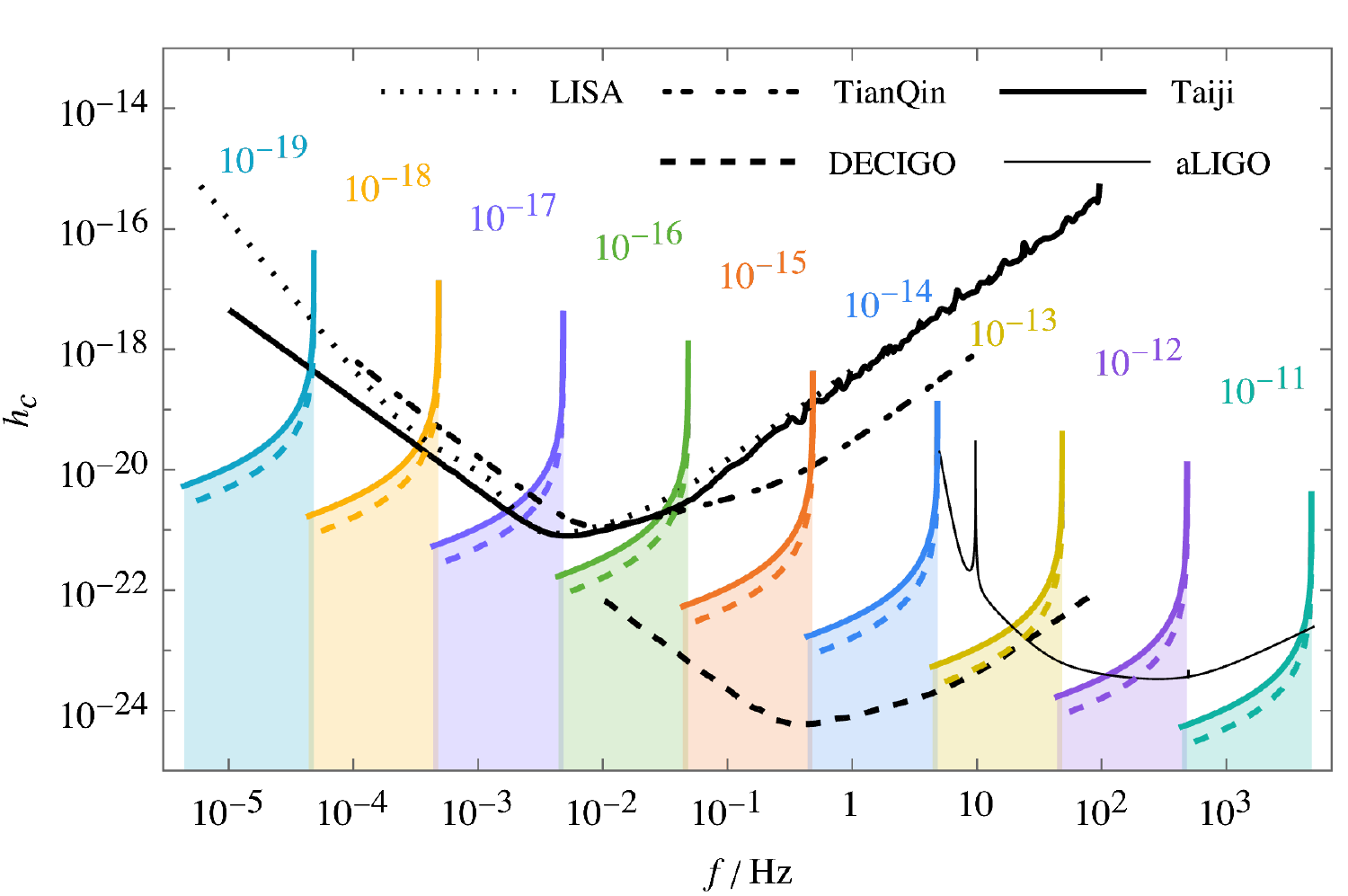}
	\caption{The dimensionless characteristic strain with the largest GW emission flux (colored solid curves) and with the smallest GW emission flux (colored dashed curves), together with the characteristic noise strain (black curves) of LISA \cite{LISA:2017pwj,Robson:2018ifk}, TianQin  \cite{TianQin:2015yph}, Taiji \cite{Luo:2021Manual}, DECIGO \cite{Kawamura:2006up}, and Advanced LIGO  \cite{Barsotti:T1800044v5}. The horizontal coordinate $f$ is the GW frequency in the detector frame. Each color band represents a scalar mass labelled in the figure with units eV. From left to right on each band, the redshift $z$ decreases from 10 to 0.001. The parameters are $a_*=0.6$, $M_s/M=0.1$, $N_{111}/N_{011}=0.1$, and the observation time $T_\text{obs}=4$~yr.
}
	\label{fig:_hc_vs_f}
\end{figure}

To connect the above calculation to the observation, we further calculate the GW strain amplitude following the method in Refs.~\cite{Brito:2017wnc,Brito:2017zvb}. The GW strain amplitude measured at the detector has the form of,
\begin{equation}
	h = h_{+} F_{+}+h_{\times} F_{\times},
\end{equation}
where the pattern functions $F_+$ and $F_-$ depend on the orientation of the detector and the direction of the GW source \cite{Ruiter:2007xx,Cutler:1997ta,Rubbo:2003ap}. We choose $\langle F_+^2\rangle =\langle F_\times^2\rangle =1/5$, $\langle F_+F_\times \rangle =0$, $\langle\left|{}_s S_{\widetilde{l}\widetilde{m}}\right|^2\rangle =1/(4\pi)$. By assuming $N_{111}\ll N_{011}$ and a single $90^{\circ}$-interferometer, the characteristic strain amplitude is approximately \cite{Brito:2017zvb},
\begin{align}
\begin{split}
  h_{c}=& \dfrac{\sqrt{N_{\text {cycles }}}}{\sqrt{10\pi}\,r}\bigg\{
\dfrac{N_{011}^{2}}{{\omega^{(011)}}^2} \frac{\left|U_{\widetilde{l} 2}^{(\widetilde{\omega}_1)}\right|^{2}}{\widetilde{\omega}_{1}^4}\\
&\hspace{0cm}
 +4 \sqrt{\dfrac{N_{011}^3N_{111}}{{\omega^{(011)}}^3\omega^{(111)}}} 
\frac{\left|U_{\widetilde{l}2}^{(\widetilde{\omega}_1)}\right| \left|U_{\widetilde{l}2}^{(\widetilde{\omega}_3)}\right|}{\widetilde{\omega}_{1}^2 \widetilde{\omega}_{3}^2}
\\
&\hspace{1cm}
\times\cos\left[\widetilde{\omega}_4\left(t-r_*\right)
-\phi_{\widetilde{l}2}^{(\widetilde{\omega}_3)}+\phi_{\widetilde{l}2}^{(\widetilde{\omega}_1)}
\right] \bigg\}^{1/2},
\end{split}
\end{align}
where $N_{\text {cycles }}$ is the number of observed cycles and $r$ is the comoving distance. Other quantities are explained below Eq.~\eqref{eq:GW_Lumi}. Note that this result is in the source frame. To obtain the $h_c$ in the detector frame, corrections due to cosmological effects should be included. All quantities with dimension $[\text{mass}]^p$ in the source frame need to be multiplied by a factor of $(1+z)^p$. Specifically, the frequencies are multiplied by $(1+z)^{-1}$, and the comoving distance is replaced by the luminosity distance. The number of observed cycles is calculated with the frequencies in the detector frame.

The GW strain amplitude in the detector frame is shown in Fig.~\ref{fig:_hc_vs_f}, together with the sensitivity curve of the current and projected GW detectors. The parameters are $a_*=0.6$, $M_s^\text{tot}/M=0.1$, $N_{111}/N_{011}=0.1$, and the observation time $T_\text{obs}=4$~yr. These parameters are suitable for the BH-condensate system in the $\{1,1,1\}$ attractor phase with $M\mu \sim 0.17$. In this phase, both $M_s^\text{tot}/M$ and $N_{111}/N_{011}$ are close to their maximum values. For each scalar mass, the redshift $z$ varies from 0.001 to 10. Because of the beats, the GW emission flux varies periodically with time. In the figure, we plot separately the $h_c$ with the largest and smallest fluxes. If the GW with the smallest fluxes could be observed, the beat-like pattern should be detected as well. From Fig.~\ref{fig:_hc_vs_f}, we find that DECIGO has a very good potential for scalars with a mass between $10^{-16}$ and $10^{-13}$~eV. Advanced LIGO is sensitive for the scalars with masses from $10^{-13}$ to $10^{-11}$~eV. LISA, Taiji, and TianQin are capable of analyzing the mass range between $10^{-18}$ and $10^{-16}$~eV.

\begin{figure*}[htb]
	\includegraphics[width=0.45\textwidth]{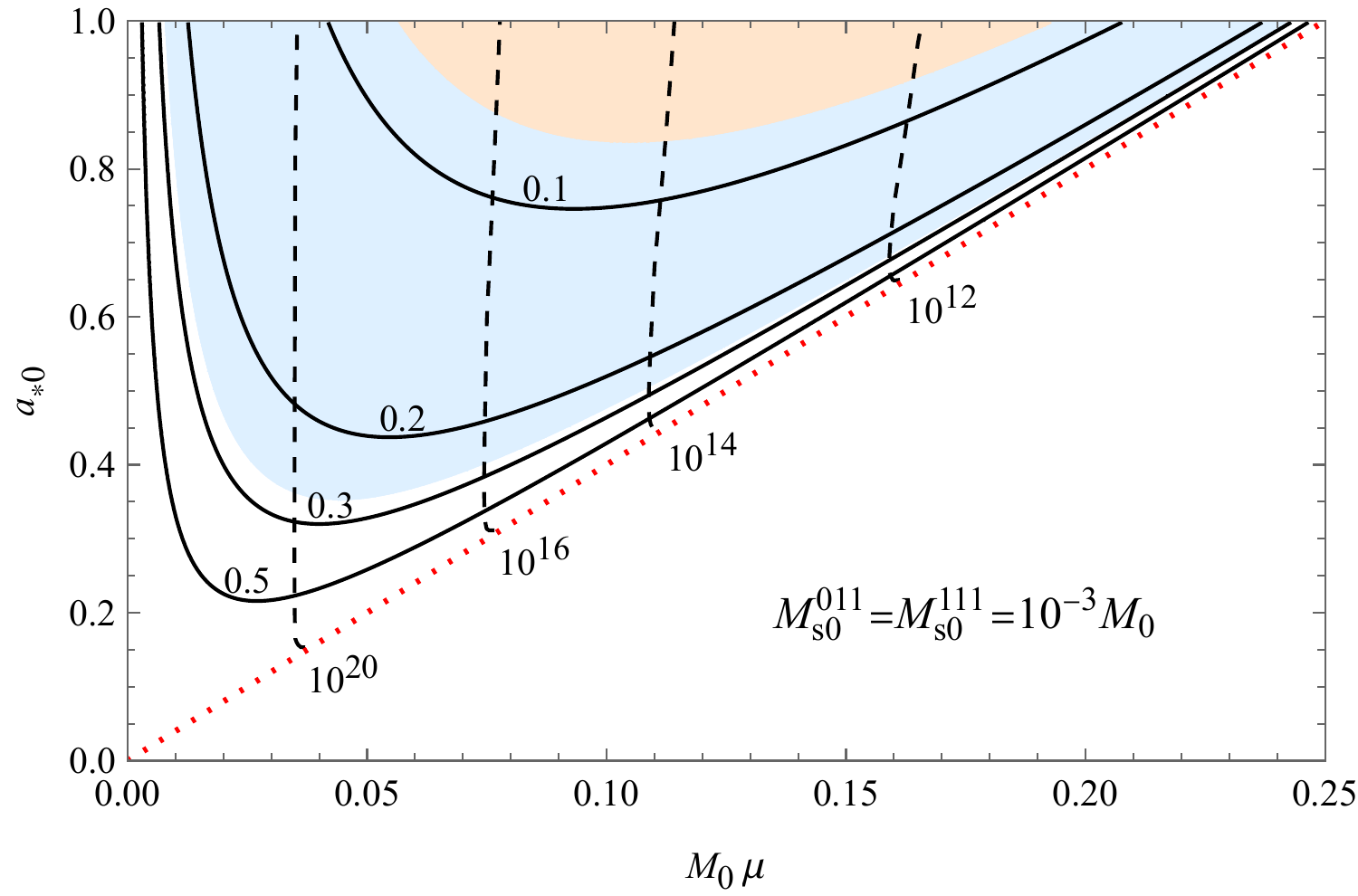}
		\includegraphics[width=0.45\textwidth]{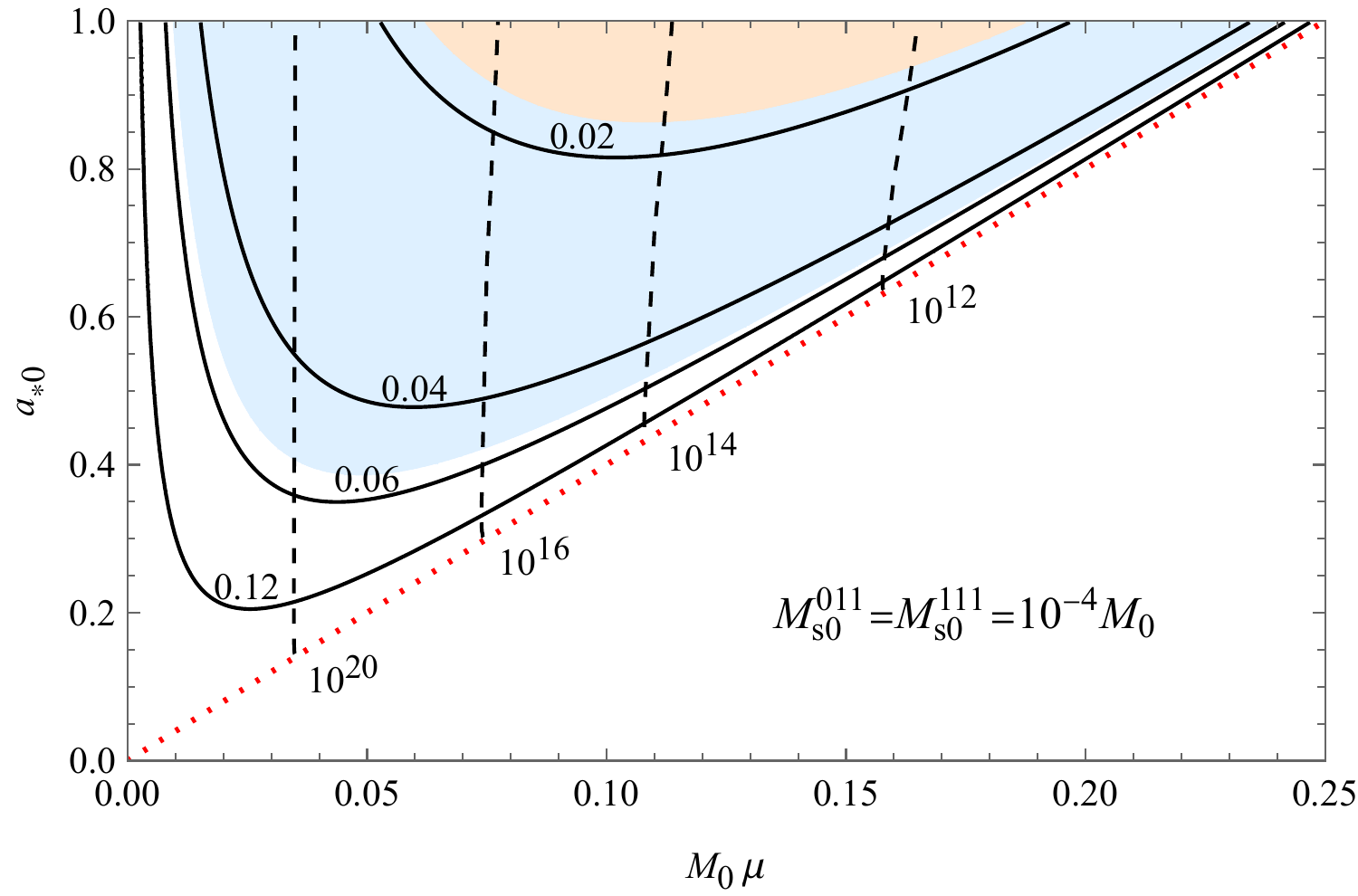}
	\caption{
The strength and the duration of the GW beat with different BH initial parameters. The initial masses of the scalar condensates are $M_{s0}^\text{011} = M_{s0}^\text{111}= 10^{-3}M_0$ (left) and $M_{s0}^\text{011} = M_{s0}^\text{111}= 10^{-4}M_0$ (right). The red dotted line is $a_{*0}=4M_0\mu$, below which the $\{0,1,1\}$ superradiance does not exist. The blue (orange) shaded region is for $M_s^\text{tot}/M > 0.01$ ($0.05$). The solid contours are calculated with  $N_{111}/N_{011}$ being the constants labelled above the curves. The dashed contours are for constant $t_2$ labelled beneath the curves, with values in the units of $M_0$. The conversion from $M_0$ to the SI units is given in Eq.~\eqref{eq:M_SI_unit}.}
	\label{fig:contour}
\end{figure*}

Here we focus on the beat-like pattern appearing at the $\{1,1,1\}$ attractor phase. The observation of the beats depends on (1) the GW emission luminosity, which is related to $M_s^\text{tot}$, (2) the size of the beat, which is determined by $N_{111}/N_{011}$, and (3) the duration of the $\{1,1,1\}$ attractor phase, which is estimated by $t_2$ in Fig.~\ref{fig:_evolution-0.1-0.99}. Given the initial masses of the $l=m=1$ condensates, the multiplication of the initial BH mass with the scalar mass $M_0\mu$, and the initial BH spin $a_{*0}$, these quantities can be reasonably estimated with the formulas given in Sec.~\ref{subsec:withoutAcc}. In Fig.~\ref{fig:contour}, we show the results with $M_{s0}^\text{(011)} = M_{s0}^\text{(111)}= 10^{-3}M_0$ and $M_{s0}^\text{(011)} = M_{s0}^\text{(111)}= 10^{-4}M_0$. When the BH-condensate system evolves into the $\{1,1,1\}$ attractor phase, the BH mass is almost unchanged, while the BH spin is approximated by $a_{*C}^{011}$ from Eq.~\eqref{eq:aStarC}. We see that quite a large initial parameter space can produce GWs with sizeable beat modulation which lasts for a long time. Interestingly, with $M_0\mu$ fixed, the region with a stronger GW luminosity has a weaker beat modulation. The golden region for observation seems to be $0.05\lesssim M_0\mu \lesssim 0.1$ and $0.5\lesssim a_{*0}\lesssim 0.7$.

Finally, this $\{1,1,1\}$ attractor phase is modified only mildly by accretion and the nonlinear effects, as illustrated in Sec.~\ref{sec:BH_evo}. In the case with accretion, the values $M_0$ and $a_{*0}$ in Fig.~\ref{fig:contour} should be interpreted as the BH mass and spin at the beginning of the decelerating phase. As a result, the results calculated above should also hold even if these effects are present.

\section{Summary}
\label{sec:_Summary}

In this work, we have carefully studied the BH-condensate evolution with scalar superradiance instability. We especially focus on the contribution from the subdominant modes with $n\geq 1$, which are usually ignored in the literature. The evolution process with these subdominant modes is much more complicated. Nonetheless, we observe that the life of each mode can be split into different phases: {\it accelerating, decelerating, attractor} and a possible {\it quasi-normal} phase depending on the accretion efficiency. With this observation, the evolution process with an arbitrary number of modes can be analyzed in a simple and straightforward way. We give explicit formulas to estimate the maximum masses of modes $\{0,1,1\}$, $\{1,1,1\}$, and $\{0,2,2\}$. Estimations of the saturation time of the $l=m=1$ modes and the depletion time of the $\{1,1,1\}$ mode are also given. Approximations of other masses and timescales could be obtained similarly. Compared to the numerical results, these approximations work reasonably well, supporting our strategy to split the life of a mode into the phases above. It also provides a simple way to analyze the BH-condensate system without solving the differential equations. 

The dominant $\{0,1,1\}$ mode and the subdominant $\{1,1,1\}$ mode coexist in the attractor phase of the latter. Due to the small difference in the periods of the two modes,  constructive interference happens when the dense regions of these two condensates overlap. On the other hand,  destructive interference happens when they are out of phase. This results in a beat-like modulation of the GW emission flux. The period of this modulation is given in Eq.~\eqref{eq:T_mod}, which is within the observation time of most projected GW detectors. We further calculate the GW luminosity and the GW strain of the beat pattern. The comparison to the noise strain of different GW detectors is shown in Fig.~\ref{fig:_hc_vs_f}. We find that DECIGO has a very good potential for scalars with a mass between $10^{-16}$ and $10^{-13}$~eV. Advanced LIGO is sensitive for the scalars with mass from $10^{-13}$ to $10^{-11}$~eV. LISA, Taiji, and TianQin are capable of analyzing the mass range between $10^{-18}$ and $10^{-16}$~eV. This GW beat pattern can be used as a signature of some BH-condensate systems, to distinguish them from other continuous sources, such as rotating neutron stars.

Below, we further summarize the main points for each of the previous sections. In Sec.~\ref{sec:_ultra-light_scalar_field_and_superradiance_modes}, we briefly review the scalar superradiant instability of Kerr BHs. We have ignored the back-reaction of the condensate as well as the self-interaction of the scalar particles. Since this topic has been well-studied in literature, we only list some important properties, especially on the subdominant modes.

In Sec.~\ref{subsec:withoutAcc}, we solve the differential equations explicitly for the BH-condensate evolution with no accretion. The adiabatic approximation is assumed, and the NLO analytic approximation of $\omega_I$ is used. We keep two dominant modes ($\{0,1,1\}$ and $\{0,2,2\}$) and two subdominant modes ($\{1,1,1\}$ and $\{1,2,2\}$). The GW emission rate of the $\{0,1,1\}$ mode in Schwarzschild spacetime has been obtained in Ref.~\cite{Brito:2014wla}. We calculate the rates of the other three modes following their method. The BH initial spin and mass are $a_{*0} = 0.99$ and $M_0=0.1/\mu$, respectively, where $\mu$ is the mass of the scalar particle. The initial masses of all four condensate modes are set as $10^{-3}M_0$. In addition to the well-known attractor of the $\{0,1,1\}$ mode, we also study the effect of the $\{1,1,1\}$ attractor. In the $\{1,1,1\}$ attractor, the energy and angular momentum of the $\{1,1,1\}$ mode is mainly absorbed by the $\{0,1,1\}$ mode. During this time, the BH evolves close to the $\{0,1,1\}$ Regge trajectory (see Fig.~\ref{fig:Attractors}). To understand the role of different modes, we further study three more scenarios: with the $\{0,1,1\} + \{0,2,2\}$ modes, with the $\{0,1,1\} + \{1,1,1\}$ modes, and with only the $\{0,1,1\}$ mode. We made several observations by comparing these scenarios, which help us to get simple formulas to estimate the important masses and timescales. The differential equations are then solved with two more initial parameter sets, and these formulas work reasonably well for the new calculations too (see Tab.~\ref{tab:comparison}).

In Sec.~\ref{subsec:withAcc}, we argue the effect of accretion based on the calculation in Ref.~\cite{Brito:2014wla}. We propose to split the life of each mode into different phases. The life of every mode has at least three phases: accelerating, decelerating, and attractor phases. The dominant mode may also have a quasi-normal phase if accretion is efficient. With this splitting strategy, the analysis of BH-condensate evolution with an arbitrary number of modes is much simplified. In Sec.~\ref{subsec:nonlinear}, we argue that the scalar self-interaction does not shut down the $\{1,1,1\}$ mode.

In Sec.~\ref{sec:GW_emission}, we study the GW emission from the scalar condensate. We first generalize the calculation framework in Ref.~\cite{Brito:2017zvb} to include the subdominant modes. Then we check our formalism by calculating the GW emission luminosity for each of the four condensate modes in Sec.~\ref{subsec:GW_single}. The obtained results are consistent with previous calculations and analytic approximations. In Sec.~\ref{subsec:_Multiple_modes}, we calculate the GW emission with the interference of the $\{0,1,1\}$ and $\{1,1,1\}$ modes. The interference term is suppressed by $\sqrt{N_{111}/N_{011}}$, where $N_{nlm}$ is the number of scalars in the $\{n,l,m\}$ mode. This mild suppression results in a sizeable beat-like pattern in the GW emission flux even when $N_{111}/N_{011}$ is small. This GW beat exists in the attractor phase of the $\{1,1,1\}$ mode. We further study the strength and duration of the GW beat pattern for different initial parameters. In a pretty large parameter space, the GW beat is strong and lasts long enough to be observed (see Fig.~\ref{fig:contour}).

\section*{Acknowledgements}
This work is supported in part by the National Natural Science Foundation of China (NSFC) under Grant No. 12075136 and the Natural Science Foundation of Shandong Province under Grant No. ZR2020MA094. This work makes use of the Black Hole Perturbation Toolkit.
\\

\appendix

\section{Tetrad Components of $\mathcal{T}_{\mu\nu}$}\label{app:Tmunu}

The three tetrad components of $\mathcal{T}_{\mu\nu}$ appear in Eq.~\eqref{eq:def_T4}:
\begin{widetext}
\begin{subequations}
  \begin{align}
	\mathcal T_{nn}(\phi_i,\phi_j)=&-\dfrac{1}{8\pi}e^{i ( m_i+ m_j) \varphi - i( \omega_i+ \omega_j)t}\rho^2\rho^{*2}\mathcal S_i(\theta)\mathcal S_j(\theta)\Big[K_i(r)\mathcal{R}_i(r)-i\Delta \mathcal{R}^\prime_i(r)\Big]\Big[K_j(r)\mathcal{R}_j(r)-i\Delta \mathcal{R}^\prime_j(r)\Big],\label{eq:_Tnn}\\
	\mathcal T_{m^*m^*}(\phi_i,\phi_j)=&\dfrac{1}{4\pi}e^{i ( m_i+ m_j) \varphi -i ( \omega_i+ \omega_j)t}\rho^2 \mathcal{R}_i(r)\mathcal{R}_j(r)\Big[A_i(\theta)\mathcal{S}_i(\theta)+\mathcal{S}_i'(\theta)\Big]\Big[A_j(\theta)\mathcal{S}_j(\theta)+\mathcal{S}_j'(\theta)\Big],\label{eq:_Tmsms}\\
	\mathcal T_{nm^*}(\phi_i,\phi_j)=&\dfrac{i}{8\sqrt{2}\pi}e^{i ( m_i+ m_j) \varphi -i ( \omega_i+ \omega_j)t}\rho^2\rho^*\bigg\{\mathcal{R}_i(r)\mathcal{S}_j(\theta)\Big[K_j(r)\mathcal{R}_j(r)-i\Delta \mathcal{R}^\prime_j(r)\Big]\Big[A_i(\theta)\mathcal{S}_i(\theta)+\mathcal{S}_i'(\theta)\Big] \notag \\ 
	&\   +\mathcal{R}_j(r)\mathcal{S}_i(\theta)\Big[K_i(r)\mathcal{R}_i(r)-i\Delta \mathcal{R}^\prime_i(r)\Big]\Big[A_j(\theta)\mathcal{S}_j(\theta)+\mathcal{S}_j'(\theta)\Big]\bigg\}
		\label{eq:_Tnms},
\end{align}
\end{subequations}
\end{widetext}
where $\mathcal{R}_i$ and $\mathcal{S}_i$ are the radial and angular functions defined in Eq.~\eqref{eq:_def_phi} and,
\begin{subequations}
\begin{align}
	A_i(\theta)=&\frac{ m_i}{\sin\theta}-a \omega_i\sin\theta,\\
	K_i(r) =& \left(r^{2}+a^{2}\right)  \omega_i- m_i a.
\end{align}
\end{subequations}
For the results with $\phi$ replaced by $\phi^*$, one could simply multiply the corresponding $m$ and $\omega$ by $-1$, and replace $\mathcal{R}$ with $\mathcal{R}^*$ in the above equations. 

\section{Calculation of $B_{\mathrm{in}}$}
\label{sec:_Solving_RH_and_Bin}

In this appendix, we solve Eq.~\eqref{eq:_radial_Teq_s=-2} with Green's function. The corresponding homogeneous equation is,
\begin{equation}
  	\begin{aligned}
  		&\Delta^{2}\frac{\partial}{\partial r}\left(\frac{1}{\Delta}\frac{\partial R_{\widetilde{l}\widetilde{m}}(r)}{\partial r}\right)+\\
  		&\left[\ensuremath{\dfrac{\widetilde{K}^{2}+4i(r-M)\widetilde{K}}{\Delta}}-{}_{-2}\lambda_{\widetilde{l}\widetilde{m}}-8i\widetilde\omega r\right]R_{\widetilde{l}\widetilde{m}}(r)=0.
	  	\end{aligned}
	  	\label{eq:_linear_radial_Teq_s=-2}
\end{equation}
\normalsize
This second-order differential equation has two asymptotic solutions at both the horizon and the infinity,
\begin{subequations}
\begin{align}
\underset{r\rightarrow r_+}{\lim}R_{\widetilde{l}\widetilde{m}}(r) \sim \Delta^{2}e^{-i\widetilde{k}r_{*}} &\  \text{or}\  e^{i\widetilde{k}r_{*}},\\
\underset{r\rightarrow\infty}{\lim}R_{\widetilde{l}\widetilde{m}}(r) \sim r^{-1}e^{-i\widetilde{\omega} r_{*}}  &\ \text{or}\ r^{3}e^{i\widetilde{\omega} r_{*}},
\end{align}
\end{subequations}
where $r_*$ is the tortoise coordinate defined in Eq.~\eqref{eq:rStar}. Two Green's functions are constructed from the asymptotic behaviors of $R_{\widetilde{l}\widetilde{m}}(r)$,
\begin{subequations}\label{eq:g_asym}
\begin{align}
	g_{\widetilde{l}\widetilde{m}}^{\infty} &\rightarrow
	\begin{cases}
	A^{\mathrm{out}}_{\widetilde{l}\widetilde{m}} e^{i\widetilde{k}r_{*}}+\Delta^{2}A^{\mathrm{in}}_{\widetilde{l}\widetilde{m}} e^{-i\widetilde{k}r_{*}} & \text{ at }r\rightarrow r_{+},\\
	r^{3}e^{i\widetilde{\omega} r_{*}} & \text{ at }r\rightarrow+\infty,
	\end{cases}\\
	g_{\widetilde{l}\widetilde{m}} &\rightarrow
	\begin{cases}
	\Delta^{2}e^{-i\widetilde{k}r_{*}} & \text{ at }r\rightarrow r_{+},\\
	r^{3}B^{\mathrm{out}}_{\widetilde{l}\widetilde{m}}e^{i\widetilde{\omega} r_{*}}+r^{-1}B^{\mathrm{in}}_{\widetilde{l}\widetilde{m}}e^{-i\widetilde{\omega} r_{*}} & \text{ at }r\rightarrow+\infty,
	\end{cases}\label{eq:RH}
\end{align}
\end{subequations}
where the coefficients $A^{\mathrm{in/out}}_{\widetilde{l}\widetilde{m}}$ and $B^{\mathrm{in/out}}_{\widetilde{l}\widetilde{m}}$ could be determined by solving Eq.~\eqref{eq:_linear_radial_Teq_s=-2} numerically. Then the solution of the inhomogeneous Eq.~\eqref{eq:_radial_Teq_s=-2} is expressed with the help of the two Green's functions,
\begin{align}
\begin{split}
R_{\widetilde{l}\widetilde{m}}&=\frac{(-1)}{W_{\widetilde{l}\widetilde{m}}}
\bigg\{ g_{\widetilde{l}\widetilde{m}}^{\infty}\int_{r_{+}}^{r}dr^{\prime}\frac{g_{\widetilde{l}\widetilde{m}}G_{\widetilde{l}\widetilde{m}}}{\Delta^{2}}\\
&\hspace{2.5cm}
+g_{\widetilde{l}\widetilde{m}}\int_{r}^{\infty}dr^{\prime}\frac{g_{\widetilde{l}\widetilde{m}}^{\infty}G_{\widetilde{l}\widetilde{m}}}{\Delta^{2}}\bigg\},\label{eq:solution_of_radial_Teq_s=-2}
\end{split}
\end{align}
where the Wronskian is defined as,
\begin{align}
W_{\widetilde{l}\widetilde{m}}= \frac{g_{\widetilde{l}\widetilde{m}}}{\Delta} \frac{dg_{\widetilde{l}\widetilde{m}}^{\infty}}{dr}-\frac{g_{\widetilde{l}\widetilde{m}}^{\infty}}{\Delta} \frac{dg_{\widetilde{l}\widetilde{m}}}{dr}.
\end{align}
At infinity, the Wronskian approaches $2i\widetilde{\omega} B^{\mathrm{in}}_{\widetilde{l}\widetilde{m}}$, and the behavior of the Green's functions are given in Eqs.~\eqref{eq:g_asym}. Then we obtain the asymptotic behavior of $R_{\widetilde{l}\widetilde{m}}$ in Eq.~\eqref{eq:R_asym_i+j}.

The value of $B^{\mathrm{in}}_{\widetilde{l}\widetilde{m}}$ needs to be calculated as the normalization. Numerically, one solves  Eq.~\eqref{eq:_linear_radial_Teq_s=-2} from the horizon with the asymptotic behavior of $g_{\widetilde{l}\widetilde{m}}$. The obtained function at infinity is then dominated by the term proportional to $B^{\mathrm{out}}_{\widetilde{l}\widetilde{m}}$. The extraction of $B^{\mathrm{in}}_{\widetilde{l}\widetilde{m}}$ is thus numerically unstable.

Below, we apply the method proposed by Press and Teukolsky in Ref.~\cite{Press:1973zz}. Eq.~\eqref{eq:_linear_radial_Teq_s=-2} can be rewritten as,
\begin{equation}
	R''\left(r\right)-AR'\left(r\right)-BR\left(r\right)=0,
\label{eq:AB_linear_radial_Teq_s=-2}
\end{equation}
where we have suppressed the subscripts for compactness. The coefficient functions are,
\begin{subequations}
\begin{align}
A & =\dfrac{2\left(r-M\right)}{\Delta},\\
B & =-\dfrac{\widetilde{K}^{2}+4i(r-M)\widetilde{K}-\left(8ir\omega +\lambda\right)\Delta}{\Delta^{2}}.
\end{align}
\end{subequations}
This equation has two asymptotic solutions at infinity \cite{Teukolsky:1973ha},
\begin{subequations}
\begin{align}
\varphi_{1}= & \,r^{3}e^{i\omega r_{*}}\left[1+\mathcal{O}\left({1}/{r}\right)\right],\\
\varphi_{2}= & \,\frac{1}{r} e^{-i\omega r_{*}}[1+\mathcal{O}(1/r)].
\end{align}
\end{subequations}
Apparently, these two solutions have very different significance at infinity.

An auxiliary function $S_1(r)$ is introduced, which satisfies (i) $S_{1}(r)\not\equiv0$ for $r> r_+$ and (ii) for $r\rightarrow\infty$. $S_{1}(r)$ and $\varphi_{1}(r)$ agree asymptotically to order $r^{1}$---i.e.,
\begin{equation}
\varphi_{1}-S_{1}=\mathcal{O}(r^0).
\end{equation}
Then one defines two new variables, 
\begin{align}\label{eq:chi_def}
\chi_i\equiv \frac{d}{dr}\left(\frac{\varphi_i}{S_1}\right)  \text{ with } i=1,2,
\end{align}
whose asymptotic behaviors at infinity are,
\begin{subequations}
\begin{align}
\chi_{1} & =\mathcal{O}(1/r^{4}),\\
\chi_{2} & =-2i\omega\dfrac{dr_{*}}{dr}e^{-2i\omega r_{*}}\mathcal{O}(1/r^{4}).
\end{align}
\end{subequations}
These two new variables have the same significance at infinity. In Ref.~\cite{Press:1973zz}, a functional form of $S_1$ is proposed, 
\begin{equation}\label{eq:S1}
S_{1}=r^{3}\exp\left(i\omega r_{*}+\dfrac{C_{1}}{r}+\dfrac{C_{2}}{r^{2}}\right).
\end{equation}
Inserting this form into Eq.~\eqref{eq:_linear_radial_Teq_s=-2} and expanding in powers of $1/r$, one could obtain the two coefficients,
\begin{subequations}
\begin{align}
C_{1} & =-\frac{2am\omega+\lambda}{2i\omega},\\
C_{2} & =\frac{6a^{2}\omega^{2}+4iamM\omega^{2}-6am\omega-\lambda+6iM\omega}{4\omega^{2}}.
\end{align}
\end{subequations}
The differential equation of $\chi_i$ is obtained from Eq.~\eqref{eq:_linear_radial_Teq_s=-2},
\begin{equation}
\chi''\left(r\right)=\mathcal{A}\chi'\left(r\right)+\mathcal{B}\chi\left(r\right)\label{eq:chi_eq},
\end{equation}
where,
\begin{align}
  \mathcal{A} &\equiv \alpha+\beta^{\prime} / \beta,\\
  \quad \mathcal{B} &\equiv \alpha^{\prime}+\beta-\alpha \beta^{\prime} / \beta,\\
  \alpha &\equiv A-\frac{2 S_{1}{ }^{\prime}}{S_{1}},\\
  \quad \beta &\equiv B+A \frac{S_{1}{ }^{\prime}}{S_{1}}-\frac{S_{1}{ }^{\prime \prime}}{S_{1}}.
\end{align}

Now, we are ready to solve for $B^{\mathrm{in}}_{\widetilde{l}\widetilde{m}}$ in Eq.~\eqref{eq:RH}. Eq.~\eqref{eq:chi_eq} is solved numerically with the boundary condition at the horizon,
\begin{align}
\lim_{r\to r_+}\chi = \frac{d}{dr}\left(\Delta^{2}e^{-i\widetilde{k}r_{*}} S_1^{-1}\right).
\end{align}
At infinity, the solution approaches, 
\begin{align}
\lim_{r\to \infty}\chi = B^{\mathrm{out}}_{\widetilde{l}\widetilde{m}}\chi_1 + B^{\mathrm{in}}_{\widetilde{l}\widetilde{m}} \chi_2.
\end{align}
Then, one can extract $B^{\mathrm{in}}_{\widetilde{l}\widetilde{m}}$ at infinity with,
\begin{align}
B^{\mathrm{in}}_{\widetilde{l}\widetilde{m}} & =\dfrac{\chi_{1}^{\prime}\chi-\chi_{1}\chi'}{\chi_{1}^{\prime}\chi_{2}-\chi_{1}\chi_{2}^{\prime}}.
\end{align}

To calculate $\chi_i$ at infinity, we define the asymptotic expressions of $\varphi_i$ as,
\begin{subequations}
\begin{align}
\varphi_1 &=r^3 \exp\left(i\omega r_* + \sum_{i=1}^\infty\frac{D_{1i}}{r^i}\right),\\
\varphi_2 &=\frac{1}{r} \exp\left(-i\omega r_* + \sum_{i=1}^\infty\frac{D_{2i}}{r^i}\right).
\end{align}
\end{subequations}
The coefficients $D_{1i}$ and $D_{2i}$ are determined by inserting these expressions into Eq.~\eqref{eq:_linear_radial_Teq_s=-2} and expanding in powers of $1/r$. By definition, there must be $D_{11}=C_1$ and $D_{12}=C_2$. Then the asymptotic expressions of $\chi_i$ are calculated from Eq.~\eqref{eq:chi_def}. In Ref.~\cite{Press:1973zz}, only the leading term is calculated for each $\chi_i$. For better numerical stability, we have calculated the first three terms, 
\begin{widetext}
\begin{equation}
\begin{aligned}\chi_{1} & =r^{-4}\left[E_{11}+\dfrac{E_{12}}{r}+\dfrac{E_{13}}{r^2}+\mathcal{O}\left(\dfrac{1}{r^{3}}\right)\right],\\
\chi_{2} & =r^{-4}e^{-2i\omega r_{*}}\left[E_{21}+\dfrac{E_{22}}{r}+\dfrac{E_{23}}{r^2}+\mathcal{O}\left(\dfrac{1}{r^{3}}\right)\right],
\end{aligned}
\end{equation}
where,
\begin{align}
	E_{11} & =-\frac{2\lambda\left(2a^{2}\omega^{2}-2am\omega+6iM\omega+1\right)+4\omega\left[2a^{3}m\omega^{2}+3a^{2}\omega+am\left(3-8M^{2}\omega^{2}\right)-3iM\right]-\lambda^{2}}{8i\omega^{3}},\\
	\begin{split}
		E_{12} & =\frac{1}{4 \omega ^4}\left\{-2 \lambda  \omega  \left(2 a^2 \omega -4 a m+5 i M\right)+\lambda ^2 (1+i M \omega)\right.\\
	&\ \ \ \ \ \ \ \ \ \ \ \left. +4 \omega ^2 \left[3 a^4 \omega ^2+2 a^3 m \omega  (-3+2 i M \omega )+3 a^2 \left(m^2+3 i M \omega \right)-i a m M \left(8 M^2 \omega ^2+9\right)-6 M^2\right], \right\} 
	\end{split}\\
	\begin{split}
		E_{13} & =-\frac{i}{16 \omega ^5}\left\{\lambda ^3-2 \lambda ^2 \left(2 a^2 \omega ^2-3 a m \omega +10 i M \omega +1\right)\right.\\
			&\ \ \ \ \ \ \ \ \ \ \ \ \ \ +4 \lambda  \omega  \left[2 a^4 \omega ^3-4 a^3 m \omega ^2+a^2 \omega  \left(2 m^2+20 i M \omega -1\right)+a m (-6-20 i M \omega )+M (-18 M \omega +5 i)\right]\\
			&\ \ \ \ \ \ \ \ \ \ \ \ \ \ +\left.8 \omega ^2 \left[2 a^5 m \omega ^3+6 a^4 \omega ^2+3 a^3 m \omega  \left(1-8 M^2 \omega ^2\right)-9 a^2 m^2+a m M \left(32 M^3 \omega ^3+15 i\right)+6 M^2\right]\right\},
	\end{split}\\
	E_{21} & =-2i\omega,\\
	E_{22} & =-4am\omega-2\lambda+4iM\omega,\\
	E_{23} & =\frac{i \left\{2 \left[a^2 \left(4 m^2+3\right) \omega ^2+a m \omega  (-1-6 i M \omega )-8 M^2 \omega ^2-9 i M \omega -6\right]+\lambda  (8 a m \omega -8 i M \omega +1)+2 \lambda ^2\right\}}{2 \omega }.
\end{align}
\end{widetext}

\section{Time-dependence of $J$ and $a_*$}\label{app:dot_a_J}

When we discuss the accelerating and decelerating phases in this article, we refer to the sign of  $\dot{a}_*$. The dimensionless BH spin $a_*$ depends on the BH spin $J$ and mass $M$. Since both of them change over time, $\dot{J}$ and $\dot{a}_*$ could have opposite signs in the presence of accretion. In this appendix, we study this possibility. 

From the energy and angular momentum conservation,
\begin{subequations}\label{eq:J_M_dot}
\begin{align}
\dot{J} &= \dot{J}_\text{ACC} - \frac{1}{\mu}\sum_{nlm} m \left(\dot{M}^{(nlm)}_s+\dot{E}^{(nlm)}_\text{GW}\right),\\
\dot{M} &= \dot{M}_\text{ACC} - \sum_{nlm}\left(\dot{M}^{(nlm)}_s+\dot{E}^{(nlm)}_\text{GW}\right),
\end{align}
\end{subequations}
where $\dot{J}_\text{ACC}  = \eta \dot{M}_\text{ACC}$. 
The factor $\eta$ depends on the ISCO radius of the Kerr BH. Here we use $r_\text{ISCO}= 6M$ as an estimate, which is the value for a Schwarzschild BH. The obtained value of $\eta$ is $3\sqrt{6}M/2$. Below, we ignore the GW emission, which does not change the conclusion qualitatively.

Using the definition $a_*=J/M^2$, one could obtain,
\begin{align}\label{eq:a_dot_1}
\dot{a}_* = \left(\frac{\eta}{M}-2a_*\right) \frac{\dot{M}_\text{ACC}}{M} +
\sum_{nlm}\left(2a_*-\frac{m}{\mu M}\right) \frac{\dot{M}_s^{(nlm)}}{M}. 
\end{align}
In the initial accelerating phase for all modes, the mass of the condensate is so small that the second term on the RHS is negligible. The relation of $\dot{a}_*$ and $\dot{J}$ is approximately,
\begin{align}
\dot{a}_* \approx \left(\frac{\eta}{M}-2a_*\right) \frac{\dot{J}}{\eta M}.
\end{align}
Thus, both $a_*$ and $J$ increase in this phase.

As the BH-condensate evolves, the $l=m=1$ modes become important, while other modes are still negligible. During this time, the $a_*$ must be larger than $a_{*C}^{011}$, suggesting $M\mu<a_*/4$. Keeping only the $l=m=1$ modes, Eq.~\eqref{eq:a_dot_1} can be transformed to,
\begin{align}\label{eq:a_dot_2}
\dot{a}_* = 2a_*\left(\mu\eta -1 \right)\frac{\dot{M}_\text{ACC}}{M} + \left(1-2\mu M a_*\right)\frac{\dot{J}}{M^2}.
\end{align}
At the transition of the accelerating phase and the decelerating phase, $\dot{a}_*$ equals zero, which implies,
\begin{align}
\dot{J}=\frac{2 a_* (1-\mu\eta)M\dot{M}_\text{ACC}}{1-2a_* \mu M}.
\end{align}
The RHS is positive, implying $\dot{J}>0$ at the beginning of the decelerating phase.

As the system further evolves in the decelerating phase, $\dot{J}$ turns to be negative, as shown in Fig.~\ref{fig:_evolution-0.1-0.99}. Then in the attractor phases of the $l=m=1$ modes, the value of $a_*$ is roughly $4M\mu$. Inserting it into Eq.~\eqref{eq:a_dot_2}, one could relate $\dot{J}$ and $\dot{M}_\text{ACC}$,
\begin{align}
\dot{J} = \frac{12\mu M^2(1-\mu\eta)\dot{M}_\text{ACC}}{1-12(M\mu)^2},
\end{align}
which is always positive. This result works, except for the end of the $\{0,1,1\}$ attractor phase. There are two possibilities. If the BH mass grows rapidly, the $\{0,1,1\}$ mode could enter the quasi-normal phase, where this mode quickly returns the energy and angular momentum to the BH, causing a positive $\dot{J}$. On the other hand, if the BH mass grows slowly, the $\{0,1,1\}$ mode is dissipated in its attractor phase. At the end of this phase, the $l=m=2$ modes cannot be ignored any longer. The BH deviates from the $\{0,1,1\}$ Regge trajectory and gradually merges to the $\{0,2,2\}$ Regge trajectory. The BH mass and spin satisfy $a_*/4 \leq M\mu <a_*/2$. Keeping both $M_s^{(011)}$ and $M_s^{(n22)}$ in Eq.~\eqref{eq:a_dot_1} and using Eqs.~\eqref{eq:J_M_dot}, we arrive at,
\begin{align}
\dot{a}_*=a_*(\mu\eta-2) \frac{\dot{M}_\text{ACC}}{M} + a_*\frac{\dot{M}_s^{(011)}}{M} + (1-\mu M a_*)\frac{\dot{J}}{M^2}.
\end{align}
At the end of the $\{0,1,1\}$ attractor phase, the first two terms on the RHS are negative. Thus, $\dot{a}_*$ turns from positive to negative earlier than $\dot{J}$.

As a summary of this appendix, we conclude that the BH angular momentum $J$ does not keep pace with the dimensionless spin $a_*$ all the time. Mathematically, it means that $\dot{J}$ and $\dot{a}_*$ do not always have the same sign. When $\dot{a}_*$ is positive, the value of $\dot{J}$ must be positive. In the time range when $\dot{a}_*$ is less than zero, the $\dot{J}$ is negative for most of the time, but $\dot{J}$ is positive at the beginning and the end of this time range.


\end{document}